\documentclass[useAMS,usenatbib,usegraphicx]{mn2e}

\usepackage{color}
\newcommand{\red}{\textcolor{black}}

\newcommand{\apj}{ApJ}
\newcommand{\apjs}{ApJS}
\newcommand{\apjl}{ApJL}
\newcommand{\pasj}{PASJ}
\newcommand{\mnras}{MNRAS}
\newcommand{\aapr}{The Astronomy and Astrophysics Review}
\newcommand{\nat}{Nature}
\newcommand{\actaa}{Acta Astron.}
\newcommand{\apss}{Astrophysics and Space Science}
\newcommand{\aap}{Astronomy and Astrophysics}

\title[Short bursts from the Magnetar 1E~1547.0$-$5408]{Spectral Comparison of Weak Short Bursts \\
to the Persistent X-rays from \\ the Magnetar 1E~1547.0$-$5408 in its 2009 Outburst}
\author[T.Enoto et al.]
{T. Enoto$^{1,2}$\thanks{E-mail: enoto@stanford.edu},
Y. E. Nakagawa$^{3}$,
T. Sakamoto$^{2}$,
and K. Makishima$^{1,4}$ \\
$^{1}$
High Energy Astrophysics Laboratory,
Institute of Physical and Chemical Research (RIKEN),
Wako, Saitama, 351-0198, Japan \\
$^{2}$ Goddard Space Flight Center, NASA, Greenbelt, Maryland, 20771, USA \\
$^{3}$ Research Institute for Science and Engineering, Waseda University,  
17 Kikui-cho, Shinjuku-ku, Tokyo 162-0044, Japan \\
$^{4}$ Department of Physics, University of Tokyo,
7-3-1 Hongo, Bunkyo-ku, Tokyo, 113-0033, Japan \\
}

\begin{document}

\date{12 September 2010}

\pagerange{\pageref{firstpage}--\pageref{lastpage}} \pubyear{2011}

\maketitle

\label{firstpage}

\begin{abstract}
In January 2009,
	the 2.1-sec anomalous X-ray pulsar 1E~1547.0$-$5408 evoked 
	intense burst activity.
A follow-up {\it Suzaku} observation on January 28
	recorded enhanced persistent emission both in soft and hard X-rays  \citep{Enoto 2010PASJ}.	
Through re-analysis of the same {\it Suzaku} data,
	18 short bursts 
	were identified 
	in the X-ray events
	recorded by the Hard X-ray Detector (HXD) and the X-ray Imaging Spectrometer (XIS).
Their spectral peaks appear in the HXD-PIN band, and 
	their 10--70 keV X-ray fluences 
	range from $\sim 2\times 10^{-9}$ erg cm$^{-2}$ to $10^{-7}$ erg cm$^{-2}$.
Thus,
	the 18 events define a significantly weaker burst sample
	than was ever obtained, 
	$\sim$$10^{-8}$--$10^{-4}$ erg cm$^{-2}$.
In the $\sim$0.8 to $\sim$300 keV band,
	the spectra of the three brightest bursts 
	can be represented successfully by a two-blackbody model,
	or a few alternative ones.   
A spectrum constructed by stacking 13 weaker short bursts
	with fluences in the range (0.2--2)$\times 10^{-8}$ erg s$^{-1}$ is less curved,
	and its ratio to the persistent emission spectrum
	becomes constant at $\sim$170 above $\sim$8 keV.
As a result, 
	the two-blackbody model was able to reproduce the stacked weaker-burst spectrum
	only after adding a power-law model,
	of which the photon index is fixed at  1.54 as measured is the persistent spectrum.
These results imply a possibility that 
	the spectrum composition employing 
	an optically-thick component 
	and 
	a hard power-law component 
	can describe wide-band spectra of both 
	the persistent and weak-burst emissions,	
	despite a difference of their fluxes by two orders of magnitude.
Based on the spectral similarity,
	a possible connection between the unresolved short bursts and the persistent emission 
	is discussed.
\end{abstract}


\section{Introduction}
\label{Introduction}

Magnetars are a peculiar subclass of isolated neutron stars 
	\citep{Duncan_Thompson_1992ApJ, Thompson_Duncan_1995MNRAS}
	with evidence for ultra-strong magnetic fields, 
	mainly emitting in the X-ray frequency.
Located along the Galactic plane and in the Magellanic Clouds,
	$\sim$9 soft gamma repeaters (SGRs)
	and 
	$\sim$12 anomalous X-ray pulsars (AXPs)
	are currently believed to be of  the magnetar class
	(for reviews, see \citealt{Woods_Thompson_2006csxs.book, Mereghetti2008A&ARv}).
X-ray pulsations have been observed from them in a period range of $P=2$--$12$~s,
	indicating slowly rotating isolated pulsars.
Together with their large period derivatives $\dot{P} \sim 10^{-12}-10^{-10}$ s s$^{-1}$, 
	the magnetic-dipole-radiation approximation indicates 
	these objects to have 
	ultra-strong magnetic fields of $B = 3.2\times 10^{19} \sqrt{P \dot{P}} {\rm \ G} \ga 3.3\times10^{13}$ G.
Thus, it is widely believed that 
	the peculiar characteristics of SGRs and AXPs 
	originate from their extremely strong magnetic fields.

The X-ray radiation from magnetars 
	emerges on a large variety of time scales and intensities.
One form is  ``persistent" X-ray emission 
	with a typical luminosity of $L_{\rm x} \sim 10^{35}$ erg s$^{-1}$,
	which is usually stable over long periods of time ($\sim$ a few month or longer).
A typical persistent X-ray luminosity of magnetars 
	exceeds by 1--2 orders of magnitude
	that available from their rotational energy losses, $\sim 10^{33-34}$ erg s$^{-1}$.
Without evidence for mass accretion,
	the emission is therefore considered to be powered by dissipation of the magnetic energies. 
The broad-band persistent X-ray spectrum 
	is generally composed of 
	blackbody-like soft X-ray component with a temperature of $\sim$0.3--0.5 keV
	\citep{Mereghetti2008A&ARv, Rea2007ApJ, Enoto2011PASJ},
	and 
	a power-law-like hard X-ray component with a photon index of $\Gamma \sim 1$
	\citep{Kuiper2006ApJ, Hartog2008A&A, Enoto2010ApJ}.
While the soft component can be considered 
	as optically-thick thermal radiation from the stellar surface,
	the hard component,
	spanning from $\sim$10 keV to $\sim$200 keV or higher,
	is considered to emerge 
	through a different and yet unidentified process  	
	(e.g., 
	\citealt{Thompson2005ApJ, Beloborodov2007Ap&SS, Baring2007Ap&SS,Fernandez2007ApJ, Heyl2005MNRAS}).

Another form of X-ray radiation from magnetars is sporadic emission of bursts 
	with a typical duration from $\sim$0.1 seconds to a few hundred seconds.
These bursts are phenomenologically classified into three types: 
	quite rare ``giant flares" 
	with $L_{\rm x} \ga 10^{45}$ erg s$^{-1}$ lasting about a few hundred seconds
	\citep{Mazets1979Natur,Feroci 2001ApJ,Hurley2005Natur},
	``intermediate flares"
	 with $L_{\rm x} \sim 10^{42}-10^{43}$ erg s$^{-1}$ lasting a few seconds 
	 \citep{Kouveliotou2001ApJ,Olive2004ApJ,Israel2008ApJ},
	and much more frequently occurring ``short bursts"
	with $L_{\rm x} \sim 10^{38}-10^{41}$ erg  s$^{-1}$ with $\sim$0.1-sec durations
	\citep{Nakagawa2007PASJ,Israel2008ApJ}.
Thus,
	these explosive events often show luminosities exceeding the Eddington limit for 
	a neutron star of $1.4M_\odot$ ($M_\odot$ being the solar mass),
	$L_{\rm Edd} \sim 1.8\times 10^{38}$ erg s$^{-1}$,
	presumably due to suppression of  the electron scattering cross sections
	in the strong field \citep{Paczynski1992AcA}.
The mechanisms of these bursts 
	are thought to be related to 
	rearrangement of the magnetic fields invoking reconnection \citep{Lyutikov2003MNRAS},
	or 
	motion and fracturing of the neutron star crust, i.e., starquake \citep{Thompson2002ApJ}.

The persistent X-ray intensity of a magnetar sometimes increases 
	by 1--2 orders of magnitude with unpredictable timing.
Such an transient enhancement lasts typically for a few months, including its gradual decay.
They are often accompanied,
	at its early phase,
	by a burst activity,
	which can even lead to discoveries of new magnetars.	
So far, such burst-active states, or outbursts, were observed from some magnetars;
	XTE~J1810$-$197
	\citep{Gotthelf2004ApJ,Ibrahim2004ApJ,Israel2004ApJ,Gotthelf2005ApJ,Bernardini2009A&A},
	CXOU~J164710.2$-$455216 \citep{Muno2007MNRAS,Israel2007ApJ},
	SGR~0501$+$4516 \citep{Enoto2009ApJ,Rea2009MNRAS, Enoto2010ApJ...715..665E},
	and 
	1E~1547.0$-$5408 \citep{Mereghetti2009ApJ, Enoto 2010PASJ}.
\red{More recently,
	there have been accumulating reports on such activities 
	from sources with much weaker dipole fields ($\la 4.4\times 10^{13}$~G); 
	SGR~0418+5729 \citep{2010ApJ...711L...1V, 2010Sci...330..944R},
	SGR~1833$-$0832 \citep{2010ApJ...718..331G, 2011MNRAS.416..205E}, 
	Swift~J1822.3$-$1606 \citep{2012ApJ...754...27R},
	and Swift~J1834.9$-$0846 \citep{2012ApJ...748...26K}.
}

The enhanced persistent emission 
	and 
	the burst activity  
	have been simultaneously observed in many activated magnetars.
However, 
	it is not yet clear how these two emission forms are physically related with each other
	in the postulated dissipation process of the magnetic energy.
One interesting possibility is that 
	the persistent emission is 
	composed of a large number of small bursts that are not individually detectable
	\red{(e.g., \citealt{1996ApJ...473..322T, Lyutikov2003MNRAS, Nakagawa2007PhD}).}
Such a possibility has been examined 
	using a cumulative number-intensity distribution of short bursts
	\citep{Gotz2006A&A, Nakagawa2007PASJ}.
However, 
	the observational information has so far remained insufficient to evaluate this possibility,
	since the studied short bursts 
	are so bright (with fluence $\ga 10^{-7}$ ergs cm$^{-2}$) 
	and infrequent 
	that their time-averaged flux is much lower than that of the persistent emission.
Therefore,
	it is interesting 
	to examine, from observations of activated magnetars, 
	whether weaker short bursts become similar in spectral shape to 
	the persistent X-ray emission,
	as recently found in SGR~0501$+$4516 	
	\citep{Nakagawa2011PASJ}.

In the present paper,
	we studied the activated magnetar 1E~1547.0$-$5408.
Following its discovery by the {\it Einstein Observatory} \citep{Lamb1981ApJ} in 1980
	and confirmation by {\it ASCA} \citep{Sugizaki2001ApJS},
	this object was recognized as a magnetar candidate, 
	located at the center of the supernova remnant G327.24$-$013 \citep{Gelfand2007ApJ},
	based on its X-ray spectrum from {\it XMM-Newton} and {\it Chandra} observations.
This was followed by a discovery of its radio pulsations at $\sim$2.07 s 
	(PSR~J1550$-$5418; \citealt{Camilo2007ApJ}),
	and of X-ray pulsations at the same period \citep{Halpern2008ApJ}.
Together with a measured period derivative, 
	$\dot{P} \sim 2.3\times 10^{-11}$ s s$^{-1}$ \citep{Camilo2007ApJ},
	its characteristic age and surface magnetic field intensity were estimated to be 
	1.4 kyr and $2.2\times 10^{14}$ G, respectively; 
	both parameters fall in the magnetar regime.
At present,
	1E~1547.0$-$5408
	is one of the fastest rotating objects among the known mangetars.  

\begin{figure}
\begin{center}
\includegraphics[width=85mm]{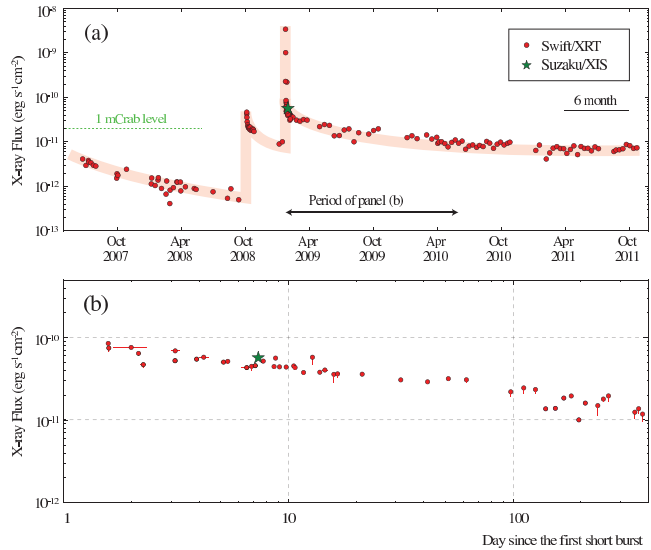}
 \caption{
Long-term X-ray monitoring of 1E 1547.0$-$5408 with the {\it Swift}/XRT,
	produced from the public {\it Swift} data archive through a standard analyses procedure,
	\red{fitted by an absorbed blackbody spectrum with the absorption left free.}	
(a) 	
The absorbed 2--10 keV X-ray flux is shown since 2007 May.
The {\it Suzaku} observation is shown in the star (green).
(b)
An expanded view of panel (a) during the 2009 outburst.
The decay is shown as a double logarithmic plot
	since the first short burst (2009-01-22 01:32:41; \citealt{Gronwall2009GCN}).
 }
  \label{fig:1e1547_2009_lightcurve.eps}
\end{center}  
\end{figure}

Figure~\ref{fig:1e1547_2009_lightcurve.eps} shows a long-term monitoring  
	of the 2--10 keV persistent emission from 1E~1547.0$-$5408.
In 2007, 
	a small enhancement was observed in its persistent luminosity \citep{Halpern2008ApJ}.
In 2008 October,
	short bursts 
	and 
	decaying persistent soft X-rays 
	have been monitored by {\it Swift} \citep{Israel2010MNRAS}.
Three month later,
	a much stronger burst activity, 
	to be dealt with in the present paper, 
	was detected on 2009 January 22 (UT).
A series of intense short bursts were recorded
	by several satellites;
	{\it Swift} \citep{Gronwall2009GCN},
	{\it INTEGRAL} \citep{Savchenko2009GCN},
	{\it Suzaku} \citep{Terada2009GCN},
	the {\it Fermi}/GBM \citep{Connaughton2009GCN, Kienlin2009GCN},
	{\it RHESSI} \citep{Bellm2009GCN},
	and 
	Konus-{\it Wind} \citep{Golenetskii2009GCN}.
Some bursts were very bright with fluences above $\sim 10^{-5}$ erg cm$^{-2}$ \citep{Mereghetti2009ApJ},
	and 
	a 150-s-long enhanced quasi-persistent emission
	was also recorded on one occasion by the {\it Fermi}/GBM \citep{Kaneko2010arXiv}.
Based on these SGR-like activities,
	1E~1547.0$-$5408 was also named SGR~J1550$-$5418 \citep{Kouveliotou 2009GCN}.	

\begin{figure}
\begin{center}
\includegraphics[width=85mm]{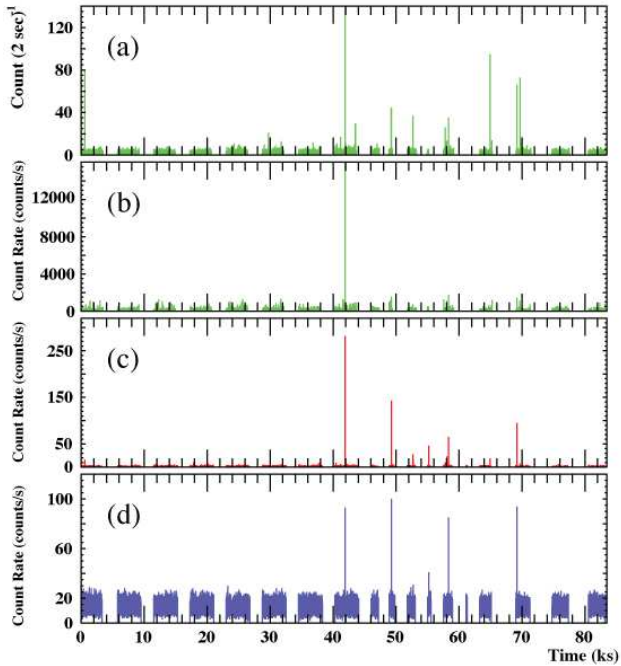}
 \caption{
Background-inclusive count rates of 1E~1547.0$-$5408 
	during the present {\it Suzaku} observation 
 	performed on 2009 January 28.
 From top to bottom,
 	panels refer to those obtained with
	XIS1 plus XIS3 (2--10 keV),
	XIS0   (2--10 keV),
	HXD-PIN (10--70 keV),
	and HXD-GSO (50--150 keV).
Data are binned into 1 sec, except for 2 sec of panel (a).
 }
  \label{fig:light_curves}
\end{center}  
\end{figure}

As shown in Figure~\ref{fig:1e1547_2009_lightcurve.eps},
	the early burst activity was accompanied by a clear enhancement of the persistent emission,
	which decayed on a timescale  of a few month.
In Enoto et al. (2010b; hereafter Paper I),
	we studied persistent emission of this object 
	using the {\it Suzaku} data acquired on 2009 January 28--29,
	or $\sim$7 days after the burst onset (Figure~\ref{fig:1e1547_2009_lightcurve.eps}), 
	when
	the 2--10 keV persistent flux became 
	by 1--2 orders of magnitude higher than 
         that during less active states in 2006 and 2007.
On that occasion,          
        a persistent {\it hard} X-ray component was detected for the first time
        from this source at least up to $\sim$110 keV.
The acquired broadband spectrum in  0.7--114 keV 
        was reproduced by an absorbed blackbody (BB) emission 
        with a temperature of $0.65\pm0.02$ keV,
        plus a prominent hard power-law (PL) with a photon index of  $\Gamma_{\rm per} \sim1.5$.
The enhanced persistent emission was also studied by 
        {\it Chandra}, {\it XMM-Newton}, {\it Swift}, and {\it INTEGRAL}
        \citep{2011arXiv1102.5419B, 2011ApJ...729..131N, Kuiper 2009ATel.1921, Hartog2009ATel.1922, 2011arXiv1102.5419B}.

In this paper, 
	we present spectral studies of short bursts 
	detected with {\it Suzaku} in the 2009 observation.
The low background of {\it Suzaku}, especially in the hard X-ray band,
	allowed us to detect much weaker short bursts
	than observed so far from any magnetars. 
Unless otherwise specified,
	we show all uncertainties at the 68\% (1$\sigma$) confidence level, 
	
\section{Observation and Data Reduction}
\label{section: EXTRACTION OF SHORT BURSTS}


\subsection{Observation and Data Screening}
\label{subsection: Observation}

We utilized the same {\it Suzaku} data (OBSID 903006010) of 1E~1547.0$-$5408 
	as used in Paper I,
	although short burst events were eliminated therein. 
The data were acquired over $\sim$1 day 
	from 2009 January 28 21:34 (UT) to January 21:32 (UT). 
As described in Paper I,
	two (XIS1 and XIS3)
	among the three X-ray Imaging Spectrometer (XIS: \citealt{Koyama2007PASJ}) sensors 
	were operated 
	with a 1/4 window mode plus  burst option, 
	repeating a 0.5-sec exposure and a 1.5-s artificial dead-time.
The other sensor, XIS0, was operated in the timing mode (P-sum mode) to attain 
	a high time resolution of $\sim$7.8 ms together with one dimensional position information.
The Hard X-ray Detector (HXD: \citealt{Takahashi2007PASJ}), 
	composed of silicon PIN diodes (HXD-PIN; 10--70 keV)
	and 
	GSO scintillators (HXD-GSO; $\sim$50--600 keV),
	was operated 
	in the normal mode with 61 $\mu$s time resolution.
Effective exposures were 42.5 ks, 10.6 ks, 10.6 ks, and 33.5 ks with
	XIS0, XIS1, XIS3, and the HXD, respectively

\begin{figure}
\begin{center}
\includegraphics[width=85mm]{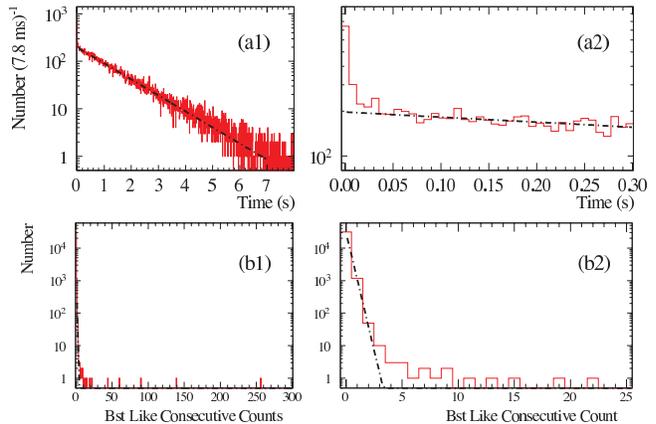}
 \caption{
(a1) A $\triangle t$ distribution of the 10--70 keV HXD-PIN events.
Dashed-dotted line indicates the best fit exponential model 
with  the average event rate of $\lambda=0.778\pm 0.001$ count s$^{-1}$.
(a2) An expanded view of panel (a1) in the 0-0.3 sec range.
(b1) Distribution of the consecutive burst-like counts 
	calculated using panel (a).
(b2) An expanded view of panel (b1).
Dashed-dotted line represents 
	the predicted chance coincidence probability of ordinary events,
	$(0.036)^{N{\rm bst}}$.
}
  \label{fig: delta-time}
\end{center}  
\end{figure}

Event screening criteria of the HXD are the same as Paper I
	except for two conditions.
We allow all geomagnetic cutoff rigidity 
	(which was $\ge$6 GV in Paper I),
        and 
        any data transfer criteria,
        in order to include as many short bursts as possible. 
These changes do not affect signal to noise ratios of
        short bursts,
        since they are almost free from the background 
        due to their short duration ($\la$1 s).
	

We utilized standard filtered files for XIS1 and XIS3,
	while 
	started our XIS0 analyses from the unfiltered event file.
These raw P-sum events were corrected and filtered 
	in a way as recommended 
	by the XIS team \citep{Matsuta2010Recipi},
	including corrections of PI values by a ftool software {\it xispi},
	selections with criteria of 
	\verb+(GRADE==0∥GRADE==1∥GRADE==2) && (STATUS >=0 && STATUS <=524287)+,
	and filtering with scripts of {\it xisrepro.xco} and {\it xis\_mkf.sel}.
We further eliminated apparent hot pixels from the XIS0 data,
	and corrected the P-sum mode timing for the known delay (31.2 ms in the present case) 
	depending on the source location on the CCD chip.

Figure~\ref{fig:light_curves} shows light curves of 1E~1547.0-5408 	
	recorded with XIS0 (2--10 keV), XIS1+XIS3 (2--10 keV), 
	HXD-PIN (10--70 keV), and HXD-GSO (50--150 keV),
	which exhibit background-inclusive average count rates of 
	3.89,
	5.77, 	
	0.76,
	and
	13.38 
	counts s$^{-1}$,
	respectively.
Several short burst events clearly appeared 
	in all light curves.
	
\subsection{Identification of Short Bursts}
\label{subsection:Delta-time distribution}

The short bursts are characterized 
	as bunched X-ray photons.
In order to identify such bunched events,
	we hence studied 
	distributions of delta-time ($\triangle t$),
	which is defined as 
	waiting times of individual X-ray photons (including background events)
	from the preceding events. 
This study utilized 
	the 10--70 keV HXD-PIN data,
	since 
	they have a better signal to noise ratio than those of HXD-GSO and XIS0,
	and the finer time-resolution (61 $\mu$s) than XIS0 ($\sim$7.8 ms).
Besides,
	unlike the XIS1 or XIS3 data with burst options,
	HXD-PIN is free from periodic data gaps.
The $\triangle$t analysis 
	provides a standard method to identify bunched events,
	and 
	is actually implemented in the HXD digital electronics 
	(HXD-DE; section 4.3 of \citealt{Takahashi2007PASJ})
	to eliminate instrumental event bunching on time scales of $\la$ a few ms.  

\begin{figure}
\begin{center}
\includegraphics[width=80mm]{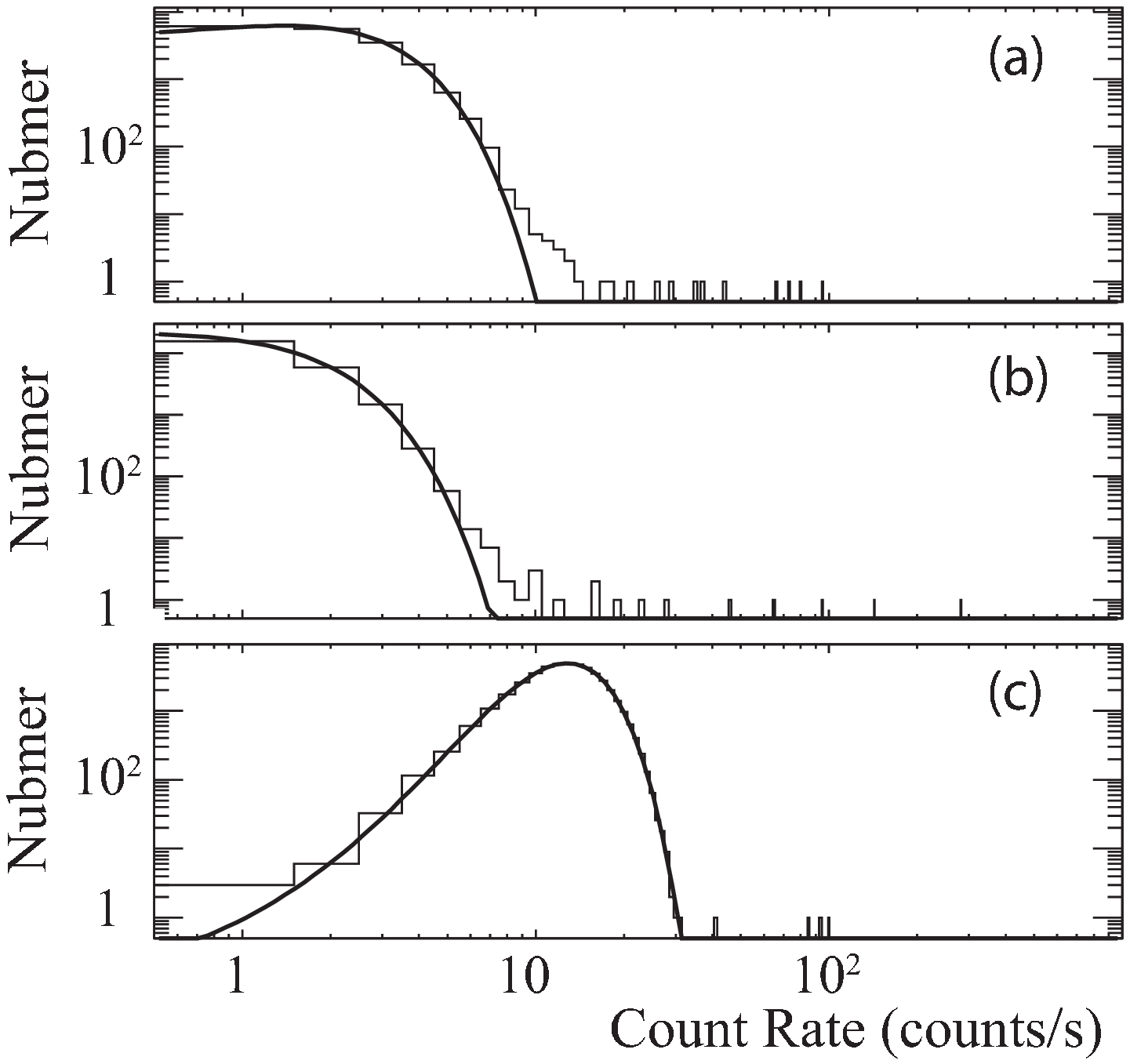}
 \caption{
 Distributions of 1-sec count rates 
 	for the XIS1+3 (panel a), HXD-PIN (panel b), and HXD-GSO (panel c) data.
Best-fit Poisson distributions are also shown.
 }
  \label{fig:lcdist}
\end{center}  
\end{figure}
	
Figure~\ref{fig: delta-time}a 
	shows 
	the calculated $\triangle t$ distribution of the 10--70 keV HXD-PIN data
	with a bin size of $t_{\rm bin}=7.8$ ms.
A large portion of these events are non X-ray backgrounds (NXB).
The $\triangle t$ values of such random events 
	follow an exponential probability distribution,
	$P(\triangle t)\propto \exp (-\lambda \triangle t)$,
	where $\lambda$ is the average event rate.
The best-fit exponential function,
	shown therein by a dash-dotted line, 
	gives $\lambda=0.778\pm 0.001$ count s$^{-1}$,
	which agrees well with the average count rate, 0.76 count s$^{-1}$,
	derived in \S\ref{subsection: Observation}.
On closer inspection	 of Figure~\ref{fig: delta-time}(a2),
	the actual $\triangle t$ distribution
	deviates 
	from the best fit model 
	below $\sim$50 ms.
This indicates that 
	some events are bunched,
	with separations much shorter than $\lambda^{-1}$.
Thus, we choose a threshold interval as 	
	$\triangle t_{\rm th} \equiv 6\ t_{\rm bin}=46.9$ ms
	to regard an event as possibly bunched.
The chance probability of observing $\triangle t \le \triangle t_{\rm th}$	
	is $\{1-\exp(-\lambda \triangle t_{\rm th})\}$=3.6\%.
	
Under the above preparation, 
	we define ``an event train of length $N_{\rm bst}$" 
	as an event bunching 
	where $N_{\rm bst}$ consecutive events are detected all with 
	a waiting time of $\triangle t < \triangle t_{\rm th}$.
Most NXB events sparsely occur
	with $\triangle t > \triangle t_{\rm th}$,
	and hence, with $N_{\rm bst}$=$0$.
Since the chance occurrence of a bunching with length $N_{\rm bst}$ is simply 
	$(0.036)^{N_{\rm bst}}$,
	we estimate those with $N_{\rm bst}=$3, 4, and 5 as 
	$4.7\times 10^{-5}$,
	$1.7\times 10^{-6}$,
	and 
	$6.0\times 10^{-8}$,
	respectively.
Considering the 10--70 keV total PIN events of $\sim 3.4\times 10^4$ counts,
	chance detections of 
	event trains with $N_{\rm bst}=$3, 4, and 5 during the entire observation
	become
	$\sim$2, $\sim$0.06, and $\sim$0.002, respectively.
Therefore,
	event trains with $N_{\rm bst} \ge 4$ would be securely identified as short bursts.
Compared to a more conventional way of detecting sudden increases 
	in the light curve,
	this method is more suited to our search for weak bursts on two points;
	it is unaffected by count binning,
	and 
	it works even when a burt has rather gradual rise (e.g., with a precursor).

Figure~\ref{fig: delta-time}b 
	shows the $N_{\rm bst}$ distribution
	in comparison with the predicted chance coincidence probability of ordinary events, $(0.036)^{N_{\rm bst}}$.
We find data points 
	with $N_{\rm bst} \ge 4$ which are significantly deviated from the fit,
	and hence regard them as short bursts.
We further merged two separate bursts into one 
	if they were detected within the adjacent 1 sec.
Using these criteria,
	we successfully detected 18 short burst events as listed in Table~\ref{tab: list of burst}.	

To confirm these detections in an independent way,
	we also studied 
	distributions of the count rates in the light curves of Figure~\ref{fig:light_curves}.
The calculated histograms are shown in Figure~\ref{fig:lcdist}.
In Figure~\ref{fig:lcdist}(b),
	the short bursts appear as significant deviations 
	from the best-fit Poisson distribution,
	which is superposed as a solid curve. 
Even only from Figure~\ref{fig:lcdist}(b), 
	fifteen among the above eighteen bursts were 
	recognized as those 1-sec bins where $\ge$8 events were detected.
Thus,
	both methods give almost the same detections.
Further corrections for the XIS1	 timing mode 
	are described in Appendix.\ref{subsection:Timing correction for the P-sum mode}.

\red{
	We also searched the Swift/BAT data for possible detections of the same short bursts. 
Using the same trigger code as used in \citet{2003AIPC..662...79G} on the 64 msec BAT data, 
	only 2 bursts were  recorded by the BAT. 
This is mainly because the BAT, with its wide field of view, has higher background.
In other words, HXD-PIN onboard {\it Suzaku} has a far higher sensitivity to weak short bursts,
	as long as the source is inside its tightly collimated field of view.
Therefore, we utilize the above {\it Suzaku} data set for the following analyses.
}

\begin{figure*}
\begin{center}
\includegraphics[width=170mm]{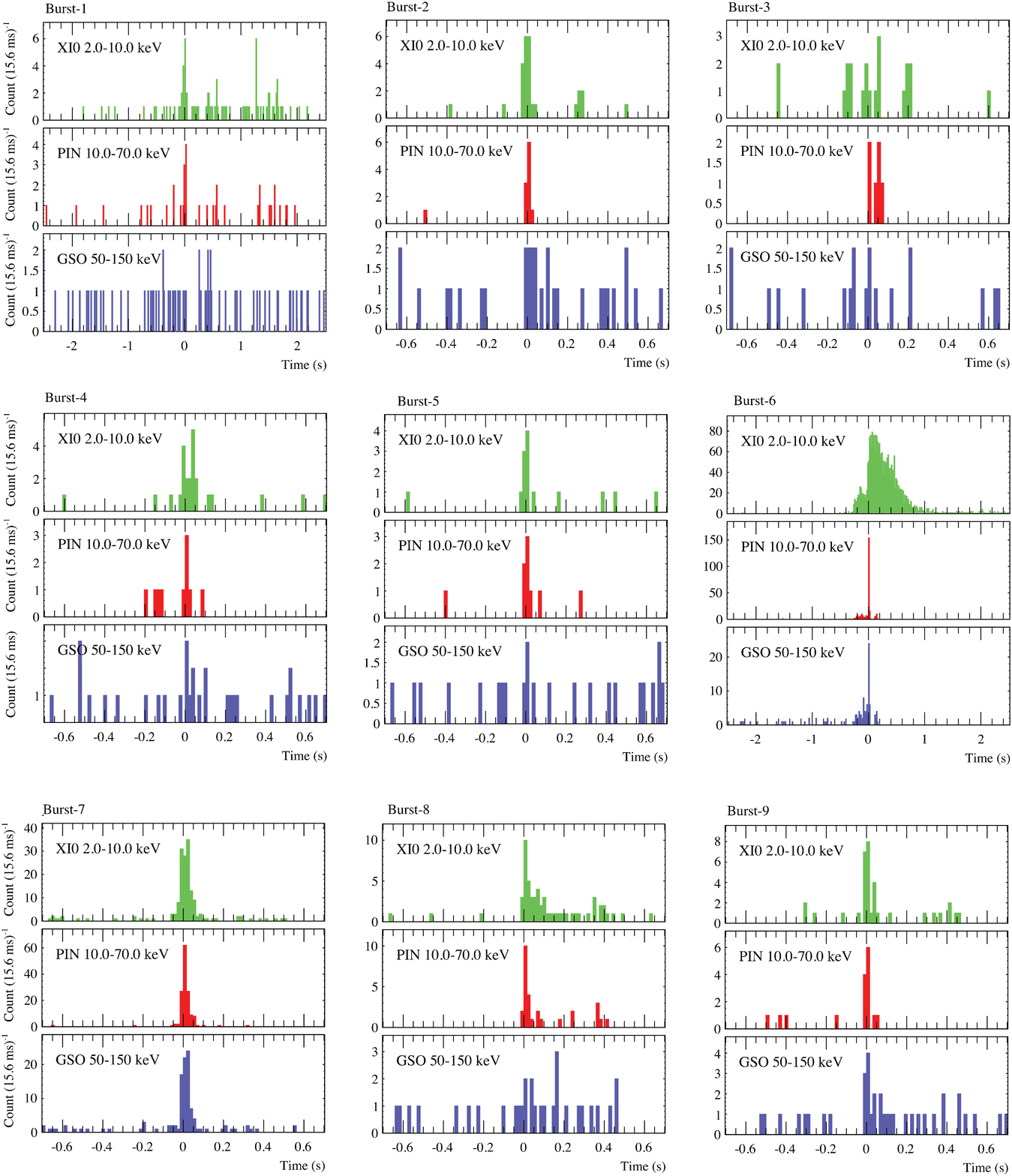}
 \vspace{-0.5cm}
 \caption{
 Light curves of the individual short burst events detected by the present {\it Suzaku} observation.
From top to bottom,
	panels refer to those obtained with XIS0, with HXD-PIN, and with HXD-GSO,
	in the 2--10, 10--70, and 50--150 keV respectively.
The time bin is 15.6 msec.
}
 \label{fig:burst_lc_1}
\end{center} 
\end{figure*}

\addtocounter{figure}{-1}
\begin{figure*}
\begin{center}
\includegraphics[width=170mm]{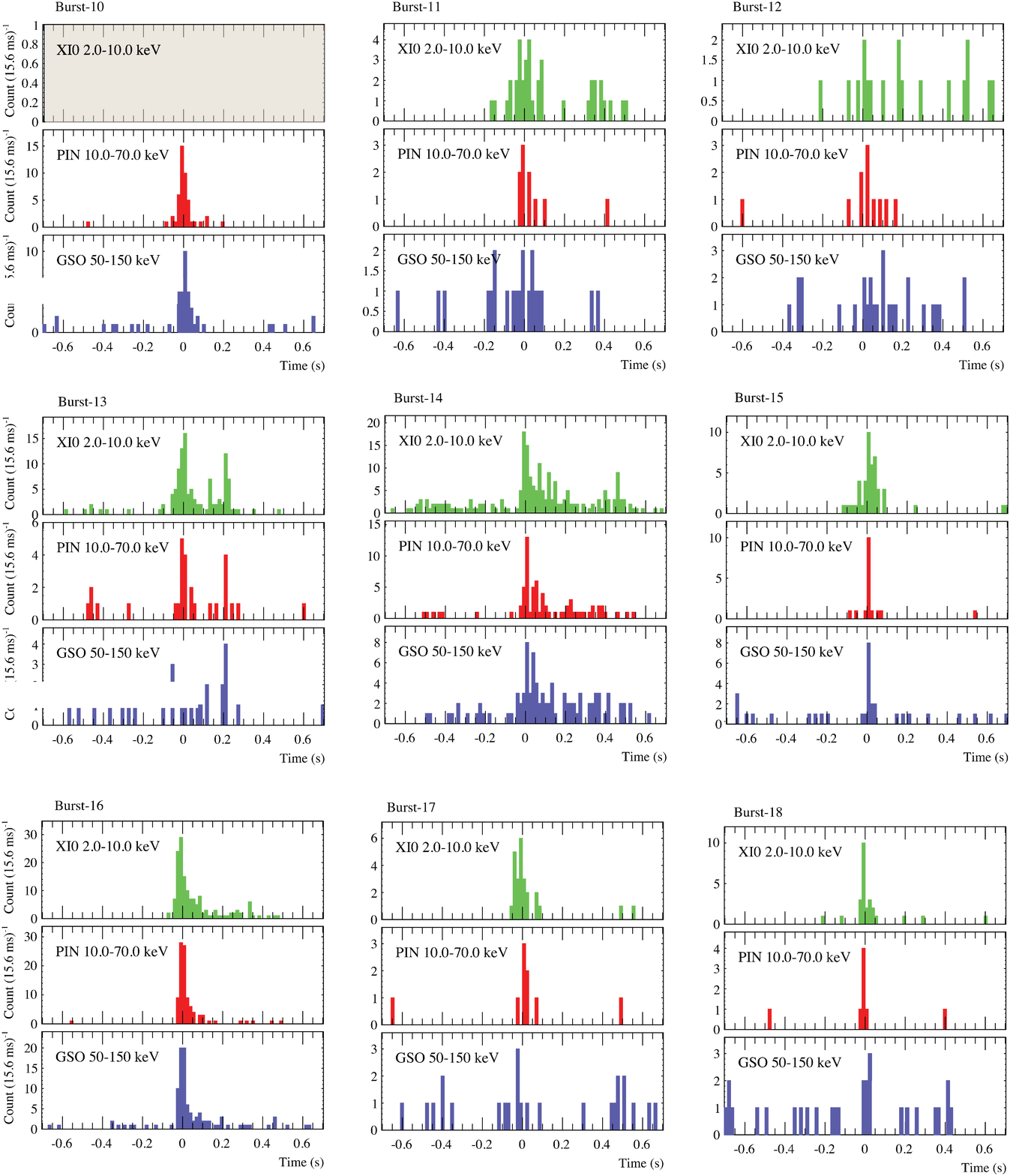}
 \vspace{-0.5cm}
 \caption{
 Continued.
 }
  \label{fig:burst_lc_2}
\end{center}
\end{figure*}

\section{ANALYSIS}
\label{section:ANALYSIS}

\subsection{Detected Bursts}
\label{subsection:Characteristics of detected bursts}

Figure~\ref{fig:burst_lc_1} 
	shows
	the XIS, HXD-PIN, and HXD-GSO light curves 
	of the bursts described in \S\ref{subsection:Delta-time distribution}.
Table~\ref{tab: list of burst} also gives 
	 their basic properties.
The  bursts are clearly detected by HXD-PIN and XIS0,
	and a considerable fraction of them by HXD-GSO as well. 
Due to the burst option,	
	the coverage of these bursts 
	with XIS1 and XIS3 (not shown in Figure~\ref{fig:burst_lc_1}) are highly incomplete. 
This however does not hamper our study,
	because we analyze below the XIS0 and HXD data of these bursts.	
We exclude Burst-6, 
	presumably the strongest one, 
	since
	a part of the HXD data was lost due to saturation of the data transfer.
Likewise, 
	Burst-10 is discarded, 
	because it fell outside the good time interval (GTI) of the XIS.
Thus,
	we hereafter analyze the remaining 16 short bursts. 
They are 
	statistically significant with chance probabilities $\la 10^{-5}$
	when combining the 1-sec data of XIS, HXD-PIN, and HXD-GSO together 
	(Appendix. \ref{subsection:The statistical significances of detected bursts}; 
	1-sec rates are also shown in Table~\ref{tab: list of burst}),
	and are free from data losses due to dead-time or pile-ups (Appendix \ref{subsection:Estimation of the data loss}).
	
\begin{figure}
\begin{center}
\includegraphics[width=60mm]{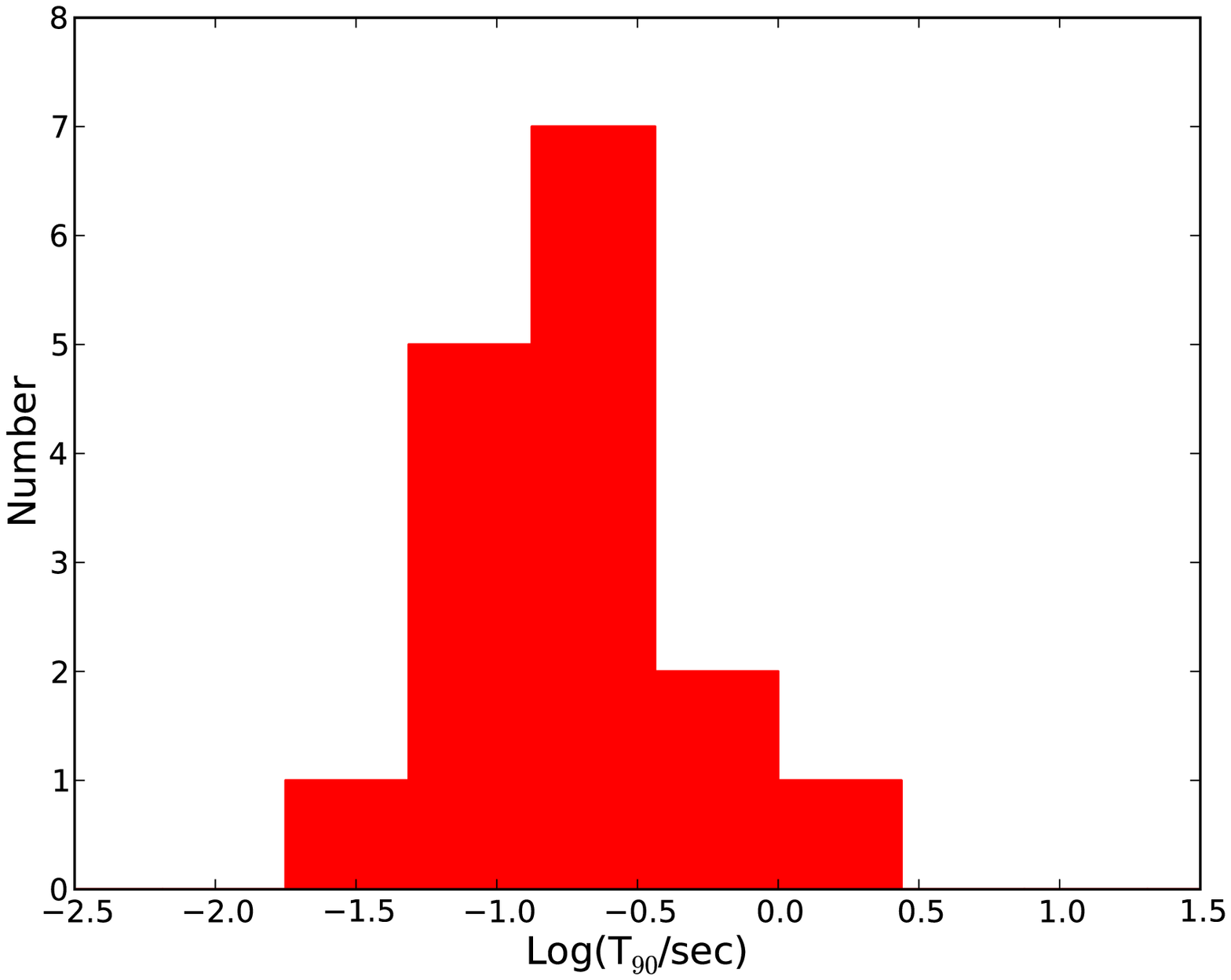}
 \caption{
Distribution of $T_{90}$ for the 16 bursts
	determined from the 10--70 keV HXD-PIN data.
 }
  \label{fig:t90}
\end{center}
\end{figure}

\begin{figure}
\begin{center}
\includegraphics[width=80mm]{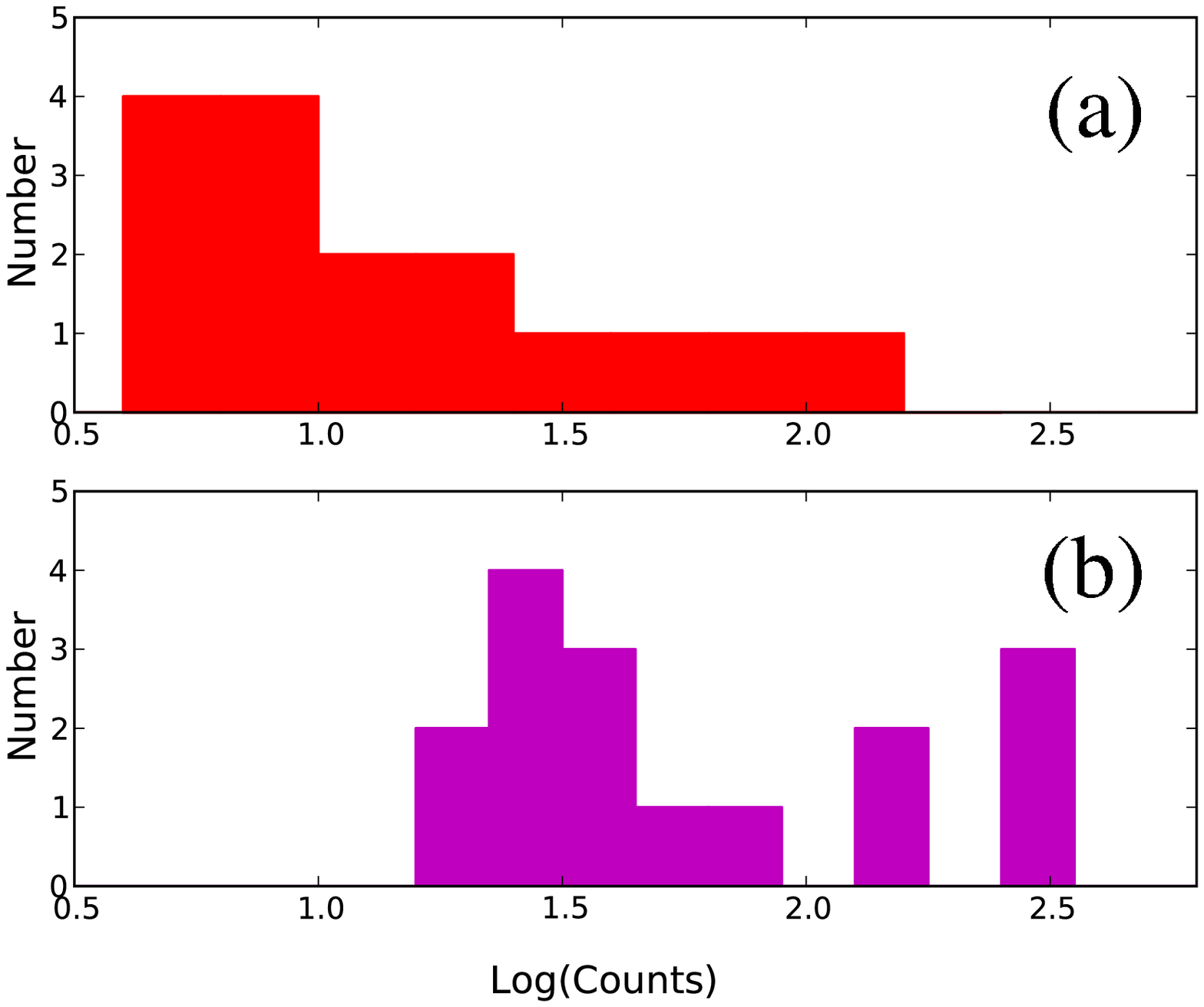}
 \caption{
(a) Distribution of burst counts recorded over $T_{90}$ in the 10--70 keV HXD-PIN data.
(b) Same as panel (a), but for the total counts 
	summed over the 2-10 keV XIS, 10--70 keV HXD-PIN,
	and 50--150 keV HXD-GSO data.
 }
  \label{fig:nt90_counts}
\end{center}  
\end{figure}

The burst peaks in Table~\ref{tab: list of burst}
	are defined as the time of the maximum count rate of HXD-PIN,
	which have higher signal-to-noise ratios than the others.
As can be seen in Figure~ \ref{fig:burst_lc_1}, 
	these peaks can be determined with a typical uncertainty of 
	a few bins, or $\la$50 ms.
Taking a closer look at individual light curves in Figure~ \ref{fig:burst_lc_1}, 
	some burst photons apparently appear 
	outside the duration of $N_{\rm bst}$ (Table~\ref{tab: list of burst}) 
	defined in \S\ref{subsection:Delta-time distribution}.
Thus, 
	using the XIS0 (2--10 keV) and HXD-PIN (10--70 keV) data co-added together, 
	we manually defined tentative burst durations 
	to cover whole possible burst X-ray photons.
Within the above durations,
	we further calculated the time period, $T_{90}$,
	starting when 5\% of the XIS0+PIN photons has been detected 
	and 
	ending when 95\% has been observed.
As shown in Figure~\ref{fig:t90} (and also Table~\ref{tab: list of burst}),
	the $T_{90}$ values range from $\sim$47 ms to 1.9 s,
	with the average of 321 ms.
\red{
This value is close to the mean values derived in recent works of the same object, 
	$T_{90}=$258 ms \citep{2012ApJ...749..122V}
	and 305 ms  \citep{2011arXiv1106.5445S},
	where the disagreement may have arisen 
	because these works utilized different sets of short bursts, 
	with different fluences and different energy ranges.   
}
Figure~\ref{fig:nt90_counts} represents
	distributions of burst photon counts $N_{\rm T90}$ detected during $T_{90}$.
Thus, $N_{\rm T90}$ of  the 10--70 keV PIN data
	are distributed from $\sim$5 to $\sim$130.
Although
	$N_{\rm T90}$ is sometimes slightly different from $N_{\rm bst}$ by up to $\sim$50\%,
	the difference is negligible in the following analyses.

For spectral analyses of the individual bursts,	
	the XIS0 and the HXD events accumulated over $T_{\rm 90}$
	were utilized as source spectra.
The background of XIS0
	was extracted using actual events on the CCD chip
	far away from the source position.	
The HXD-PIN background was not subtracted, 
	since 
	the average background rate ($\sim$0.8 Hz in 10--70 keV)
	implies at most $\sim$1 photon in $T_{\rm 90}$.
On the other hand,
	for the HXD-GSO analyses,
	we produced the background spectra from
	the simulated GSO events \citep{Fukazawa2009PASJ}
	during a 20 sec period around the corresponding burst,
	and subtracted them after scaling to the duration of $T_{90}$.
The GSO background typically contains $\sim$8 counts per bursts, 
	which amount to $\sim$30\% of the signal counts.  	
Although we did not produce background from the actual GSO data taken before/after the bursts 
	to avoid possible contamination of the persistent or burst emission from the source,
	this alternative method gives consistent results. 
We employed standard response files (epoch 5) for the HXD,
	while 
	used the response (rmf) and auxiliary response (arf) files for XIS0
	produced as in \S\ref{subsection: Observation}.

\begin{figure}
\begin{center}
\includegraphics[width=85mm]{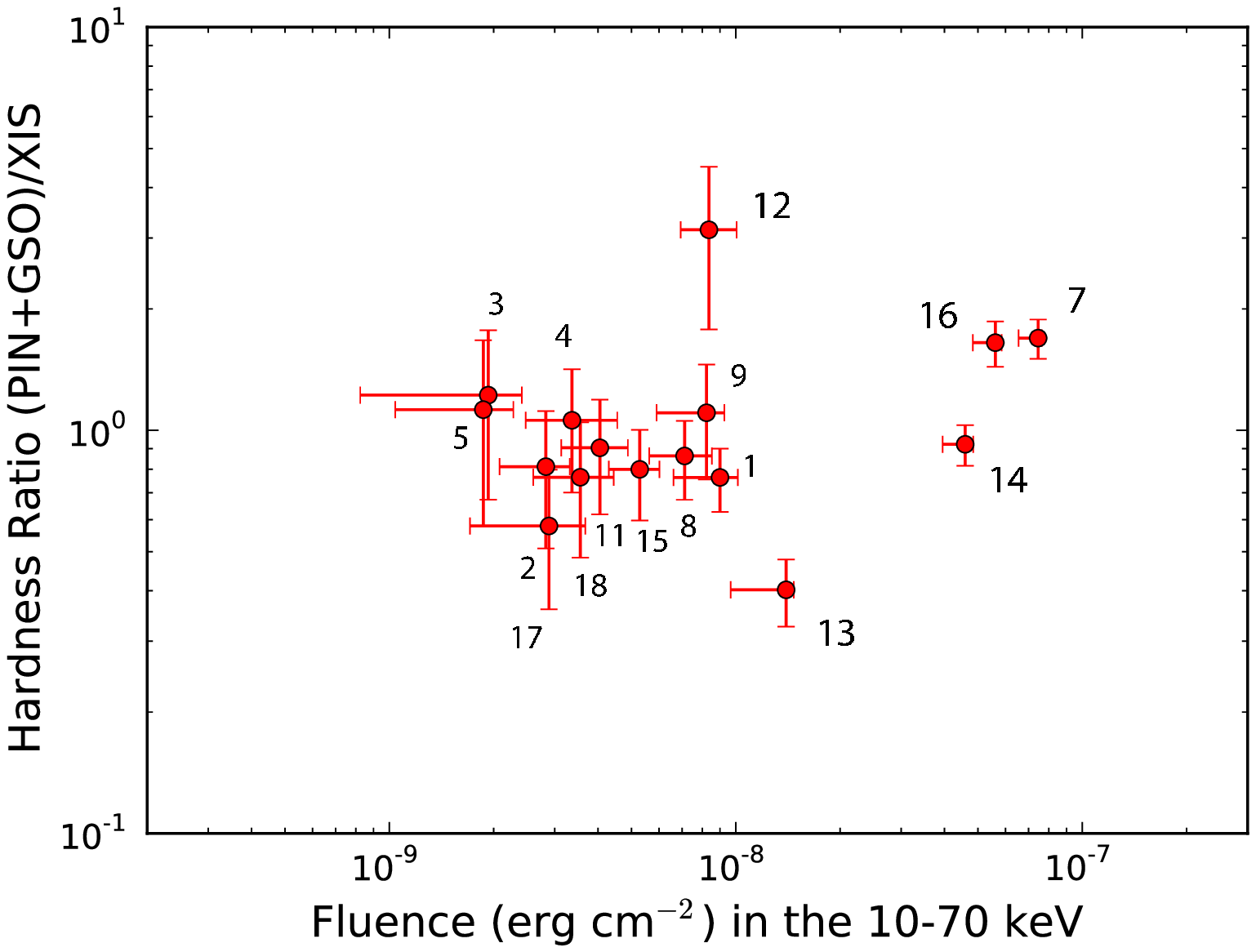}
 \caption{
Hardness ratios of the detected X-ray photons $N_{T90}$ during $T_{90}$
	calculated as (PIN+GSO)/XIS0,
 	as a function of the 10--70 keV fluence.
Numbers in the figure indicate the corresponding burst ID.
	}
  \label{fig: gamma_vs_fluence}
\end{center}  
\end{figure}

The average fluxes and fluences of these bursts
	were estimated using 
	a single PL model, 
	with a column density of the photo-absorption 
	fixed at $N_{\rm H}=3.2\times 10^{22}$ cm$^{-2}$
	after Paper I.
The resultant photon indices and the 10--70 keV X-ray fluxes
	are listed in Table~\ref{tab: list of burst}.
In some bursts,
	this PL model did not give an acceptable fit
	mainly due to a high energy signal deficiency.
For such bursts,
	we alternatively utilized 
	a cutoff power-law model (CutPL) 
	with the same fixed photo-absorption.
The 10--70 keV fluences integrated over $T_{\rm 90}$ are also shown in Table~\ref{tab: list of burst}.
As one of the characteristics of the present burst sample, 
	Figure~\ref{fig: gamma_vs_fluence} shows 
	harness ratios of the bursts photons detected by the HXD to those by the XIS, 
	as a function of their fluences. 
The fluences are distributed 
	in the range of $\sim 10^{-9}$--$10^{-7}$ erg cm$^{-2}$,
	and 
	their hardness ratios ranges from $\sim$0.4 to $\sim$3.
\red{
Although 
	a hardening trend towards weaker bursts was reported from SGRs \citep{Gogus2001ApJ},
	and a softening trend was reported from AXPs \citep{2004ApJ...607..959G},
	the present sample in Figure~\ref{fig: gamma_vs_fluence} does not show 
	any significant corrections between the hardness and fluence.
}


\begin{figure*}
\begin{center}
\includegraphics[width=160mm]{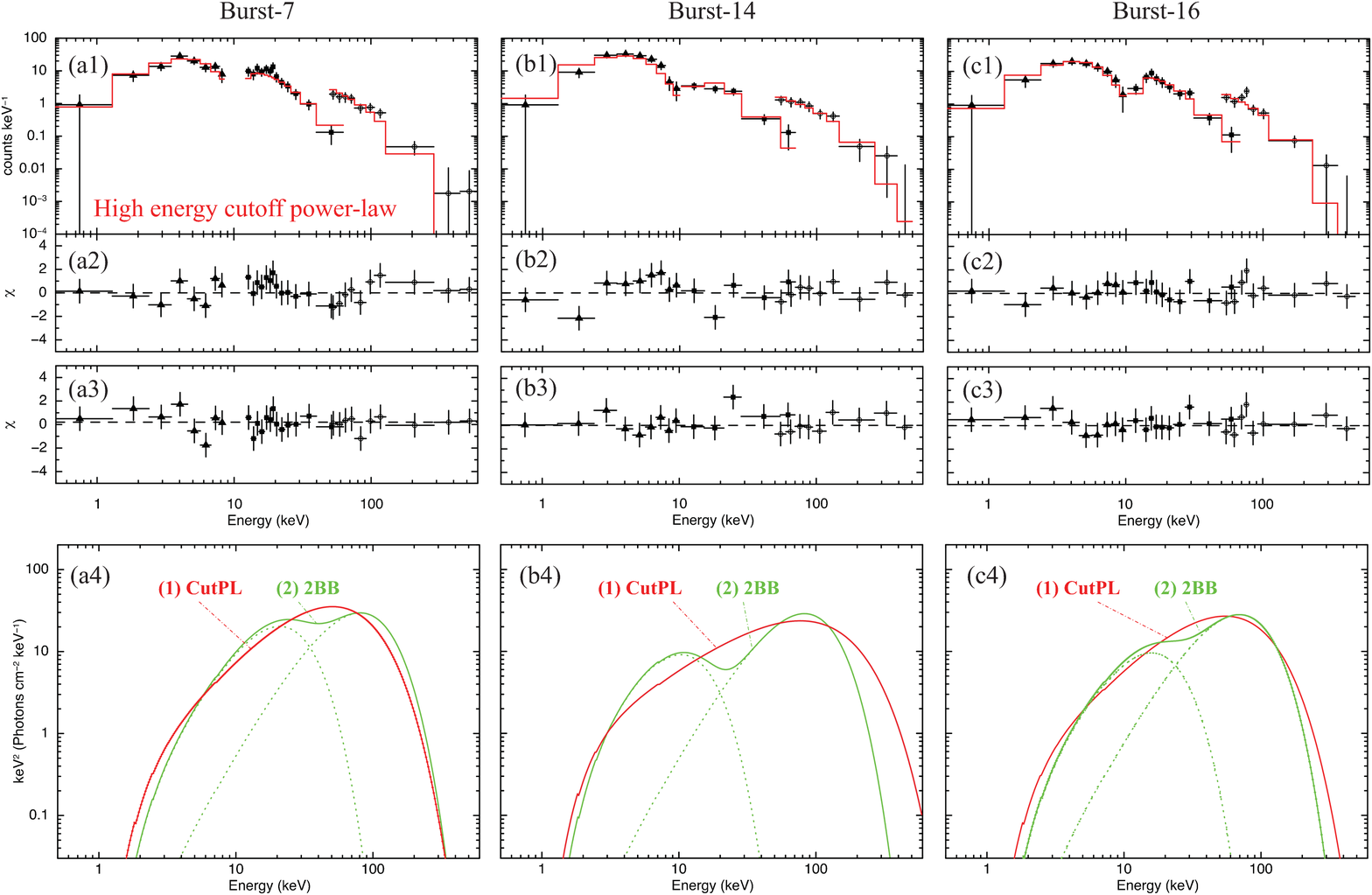}
 \caption{
X-ray spectra of the three brightest bursts in our sample;
	Burst-7 (panels a1-a5),
	Burst-14 (panels b1-b5),
	and Burst-16 (panels c1-c5).
Raw count spectra with XIS0, HXD-PIN, and HXD-GSO are shown in panel (1) with the best-fit CutPL model,
	and corresponding residuals are shown in panel (2).
Panels (3) are residuals when the data are fitted by the 2BB model.
Panels (4) are comparisons among the inferred best-fit 
	CutPL (red) and 2BB (green) models.
The spectra are shown after multiplied by individual T90 values.
The XIS0 and HXD-GSO background were subtracted as described
 in \S\ref{subsection:Characteristics of detected bursts}.
 }
 \label{fig:moderate-intensity_short_burst_spec}
\end{center} 
\end{figure*}


Among the above 16 short bursts, 
	we selected, 
	for our detailed spectral analyses,
	three outstandingly bright ones, Burst-7, 14, and 16,
	which all satisfy $N_{\rm bst}\gid 60$ 
	and show fluences above $2\times 10^{-8}$ erg s$^{-1}$ (Figure~\ref{fig: gamma_vs_fluence}).
Although these are the brightest bursts among the present sample
	(except for Burst-6 which was discarded),
	they are still weaker compared to most of the bursts used in previous studies,
	of which the fluences are typically $\sim$$10^{-8}$--$10^{-4}$ erg cm$^{-2}$.
Below we apply 
	several spectral modelings to these bursts
	in the $\sim$1--300 keV energy range.
For all models,	
	the column density for photo-absorption is fixed	
	at $N_{\rm H}=3.2\times 10^{22}$ cm$^{-2}$
	as before, unless otherwise noted.
The normalization factor of XIS0 to the HXD
	was fixed at $1.08$ based on the correction as described in \S\ref{Response correction},
	although 
	allowing it to float has an insignificant effect on the following results.
	
The single PL model with the fixed $N_{\rm H}$ gave 
	acceptable fits to none of the three bursts,
	with reduced $\chi^{2}_{\nu}$ values of 4.92, 2.83 and 3.89,
	for Burst-7, 14, and 16, respectively.
Even if we make $N_{\rm H}$ free,
	the fits were still unacceptable ($\chi^2_{\nu} \sim$2).
Discrepancies of these fits 
	originate from
	model excess in the lowest and highest spectral ends.
An optically thin bremsstrahlung model
	is not successful either ($\chi^2_{\nu} > 2$).
Then,
	we fitted these spectral data using the CutPL model,
	and 
	obtained acceptable fits as summarized 
	in Table~ \ref{tab: list of spectral fittings}
	and 
	in Figure~\ref{fig:moderate-intensity_short_burst_spec}.
The fits yielded 	
	a photon index of $\Gamma \sim 0.1$--$0.8$
	and 
	a cutoff energy of $E_{\rm cut} \sim$27--66 keV.
As alternative models, we utilized a two blackbody model (2BB).
As shown in Table~\ref{tab: list of spectral fittings}, 
	it was also successful 
	on the three bursts, 
	and yield the lower and higher temperatures of ∼3–5 keV and ∼18–20 keV, respectively. 	


In Figure~\ref{fig:moderate-intensity_short_burst_spec} (a4), (b4) and (c4),
the above two best-fit spectral models (the CutPL and 2BB)
	are compared in $\nu F_{\nu}$ forms.
They give relatively similar spectral shapes, 
	except in the higher energy range above $\sim$100 keV 
	where the models become unconstrained.
In both models,
	spectral peaks appear in the HXD-PIN band ($\sim$50--100 keV).
Employing conventionally the CutPL model, 
	absorption-corrected 0.2--300 keV fluences 
	of 
	Burst-7,
	Burst-14,
	and 
	Burst-16,
	are obtained as 
	$1.1\times 10^{-7}$, 
	$9.3\times 10^{-8}$, 
	and $8.7\times 10^{-8}$ erg cm$^{-2}$,		
	respectively.
The fluence becomes $\sim$4--8$\times 10^{-8}$ erg cm$^{-2}$
	if calculated in the 10--70 keV band. 	

In Figure \ref{fig:compare_color}a, 
	we compare the spectrum of the brightest Burst-7 
	with the persistent X-ray spectrum
	recorded during the same observation (Paper I).
While the persistent spectrum is apparently composed of two components,
	the burst spectrum is more curved,
	without apparent evidence for such two-component nature.


\subsection{Weak short bursts}
\label{subsection: Weak short bursts}

\begin{figure}
\includegraphics[width=85mm]{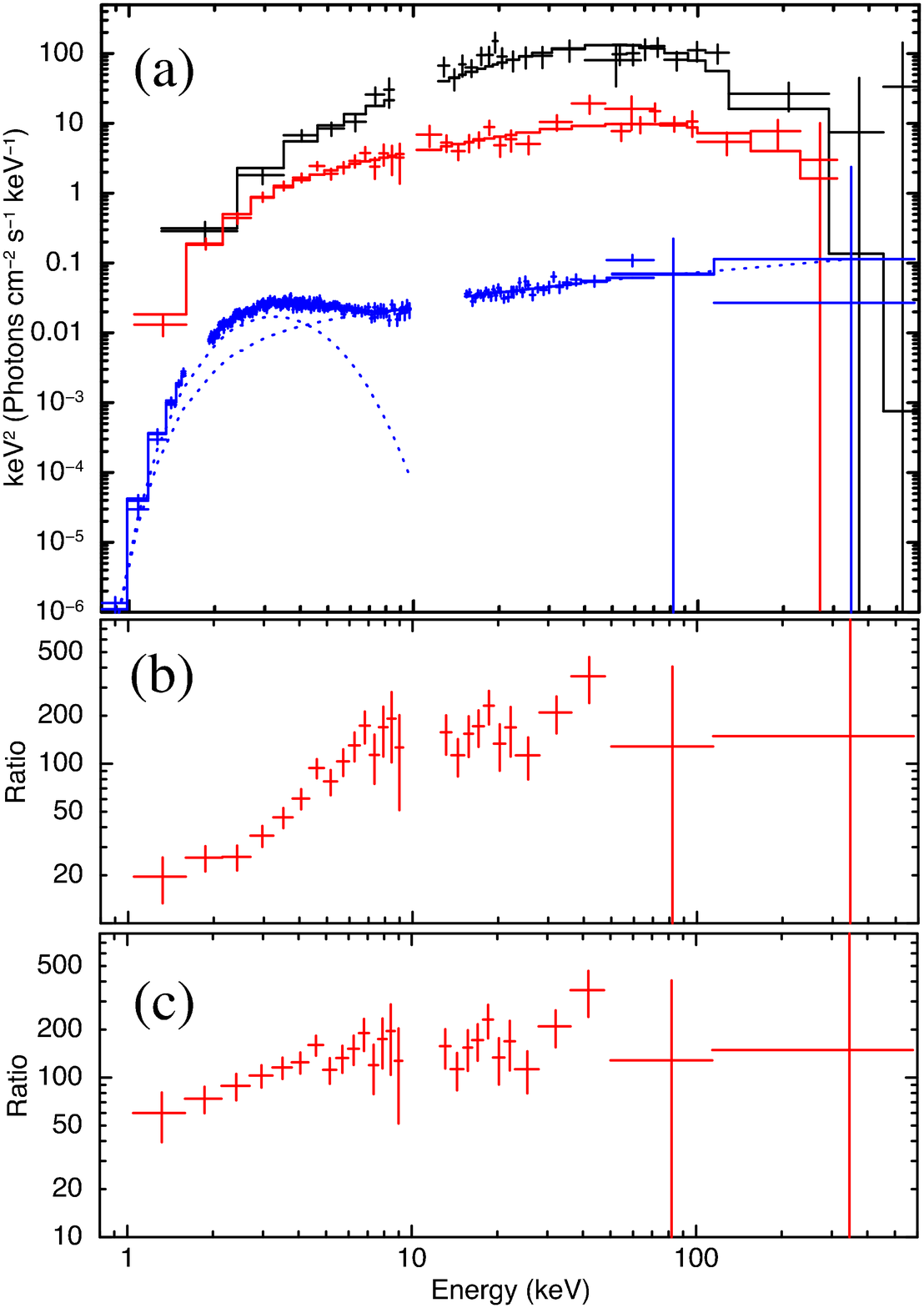}
 \caption{
	(a) Comparison of  X-ray spectra of 1E~1547.0$-$5408 in $\nu F_{\nu}$ forms.
Top (black), middle (red), and bottom (blue) spectra represent
	Burst-7,
	the accumulated weak short bursts,
	and 
	the persistent emission from Paper I, respectively.
The CutPL model was employed to deconvolve the  
	Burst-7 (Table~\ref{tab: list of spectral fittings}) 
	and the accumulated weak-burst spectra (Table~\ref{tab: list of spectral fittings of cummulative}),
	while 
	a PL plus a BB for the persistent emission.
The column density of the photo-absorption was fixed at $3.2\times 10^{22}$ cm$^{-2}$ 
	in all cases.
(b) 
The ratio between the red and blue spectra in panel (a).
(c) 
Same as panel (b),
	but after eliminating the BB component from the persistent emission.
 }
 \label{fig:compare_color}
\end{figure}

\begin{figure}
\begin{center}
\includegraphics[width=80mm]{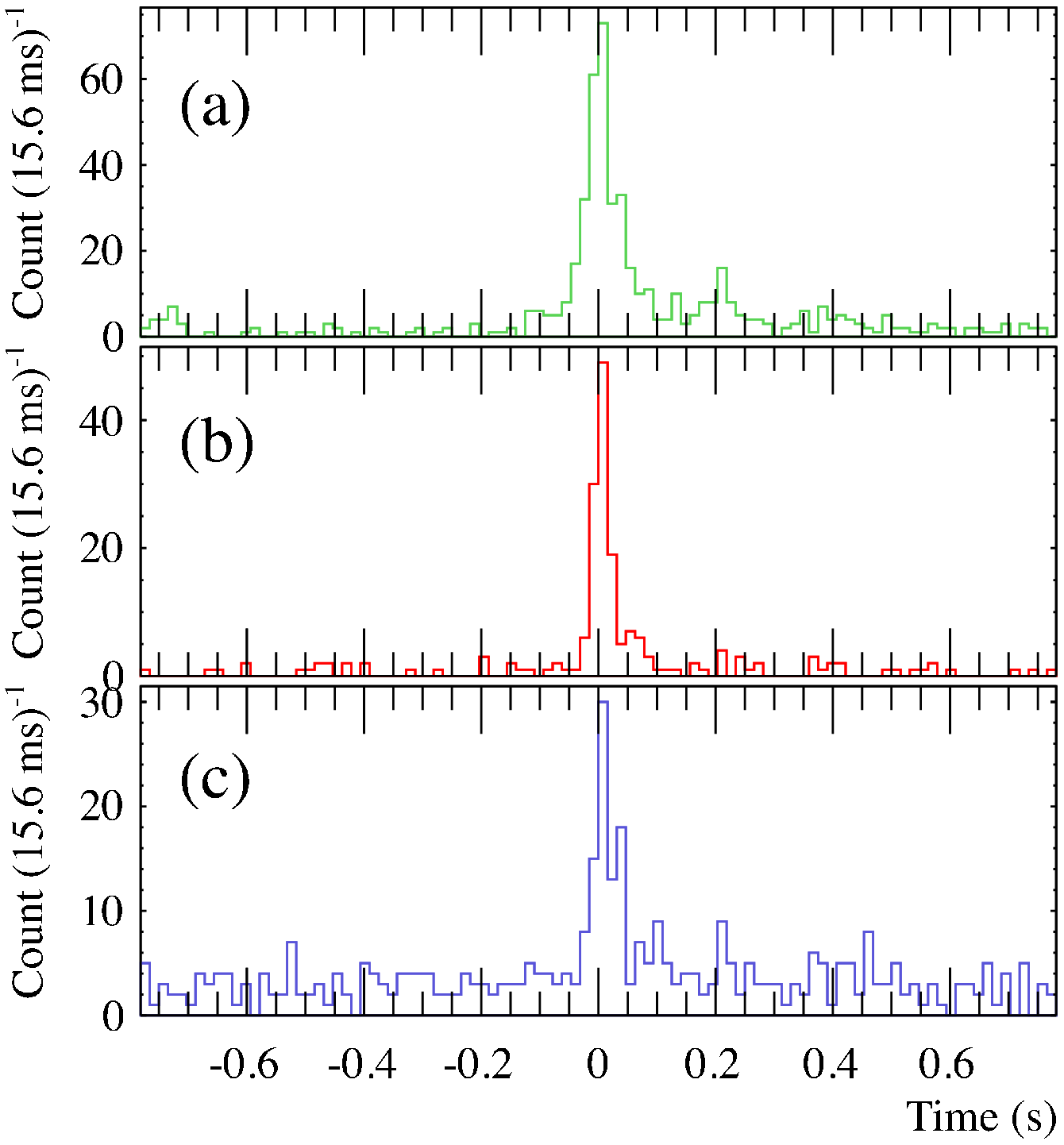}
 \vspace{3.5cm}
 \caption{
 Light curves obtained by stacking the 13 weaker short bursts, 
 	with XIS0 (panel a), 
	HXD-PIN (panel b),
	and HXD-GSO (panel c), 
	in the 2--10, 10--70, and 50--150 keV band, 
	respectively.
 }
  \label{fig:weak_stacked_lc_mod.eps}
\end{center}  
\end{figure}

As shown in Figure~\ref{fig: gamma_vs_fluence},
	the remaining 13 short bursts 
	have considerably lower 10--70 keV fluences than the three studied above,
	distributed below $2\times 10^{-8}$ erg cm$^{-2}$.
They have poorer statistics,
	and also tend to show similar harness ratios around $\sim$1.0.
Therefore, 
	we have stacked their spectra together for detailed analysis,
	with accumulated total exposure of 3.7 sec.
In order to justify the stacking procedure,
	we took spectral ratios of each burst to the stacked one, 
	to find that the ratios can be fitted in each case successfully 
	by a constant with a reduced chi-square of $\la$1.0.
Therefore, 
	the 13 bursts are concluded to have consistent spectral shapes, 	
	and hence the stacking procedure can be justified.
The derived 13 constant ratios are distributed from 0.55 to 3.3,
	with the average and standard deviation of 1.56 and 1.05, respectively.
This distribution, 
	ranging by a factor of 6,
	agrees with that of the fluecne shown in Figure \ref{fig: gamma_vs_fluence}.
The analysis here	
	utilized the same responses 
	as those of the brighter three short bursts,
	and
	in the same way as the previous analyses; 
	the background was subtracted from the XIS0 and HXD-GSO data.

\begin{figure}
\includegraphics[width=85mm]{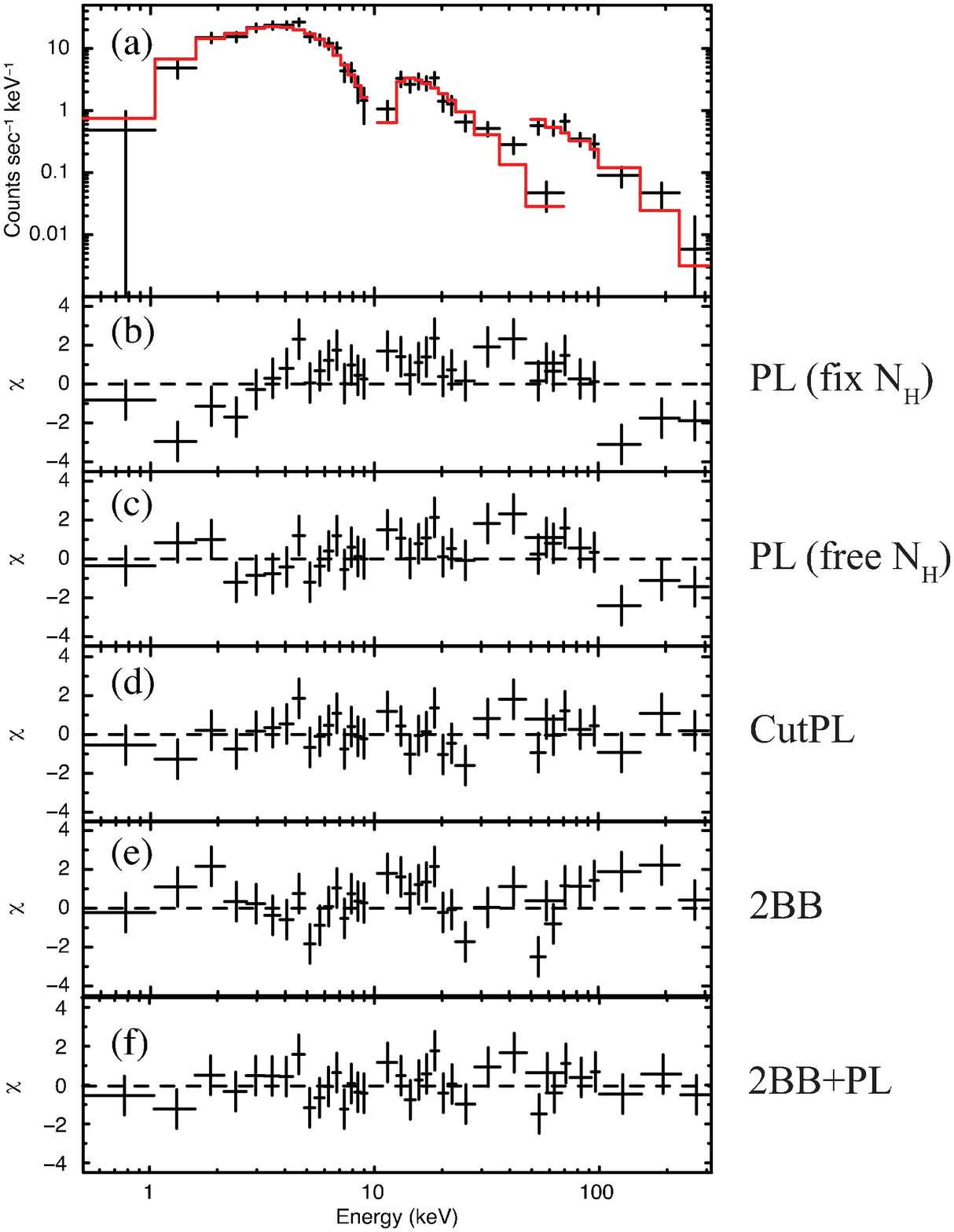}
 \caption{
(a) 
Background-subtracted and response-inclusive spectra of 
	the stacked weak bursts from 1E~1547.0$-$5408,  
	compared with the best fit  CutPL model
	(Table~\ref{tab: list of spectral fittings of cummulative}).
Panels (b)--(f) show 
	residuals when using 	
	a PL model with the fixed column density,
	a PL with the column density left free,
	a CutPL model,
	a 2BB model,
	and
	a 2BB+PL model,
	respectively.
 }
  \label{fig:accumulated}
\end{figure}

Figure~\ref{fig:weak_stacked_lc_mod.eps}	
	shows a stacked light curve of these weaker short bursts
	accumulated with reference to their peak times.
Thus, the burst emission is highly significant even in the HXD-GSO band.	
Figure~\ref{fig:accumulated}a 
	shows 
	the raw spectrum of this cumulative weak-burst data
	after the background subtraction.
Its average 10--70 keV flux 
	is 
	by an order of magnitude lower 
	than those of the three brightest ones.
The HXD-GSO background becomes comparable to the signal level around $\sim$130 keV,
	and 
	we can claim 
	the HXD-GSO detection at least up to 150 keV at 2.8$\sigma$.

\begin{figure}
\includegraphics[width=85mm]{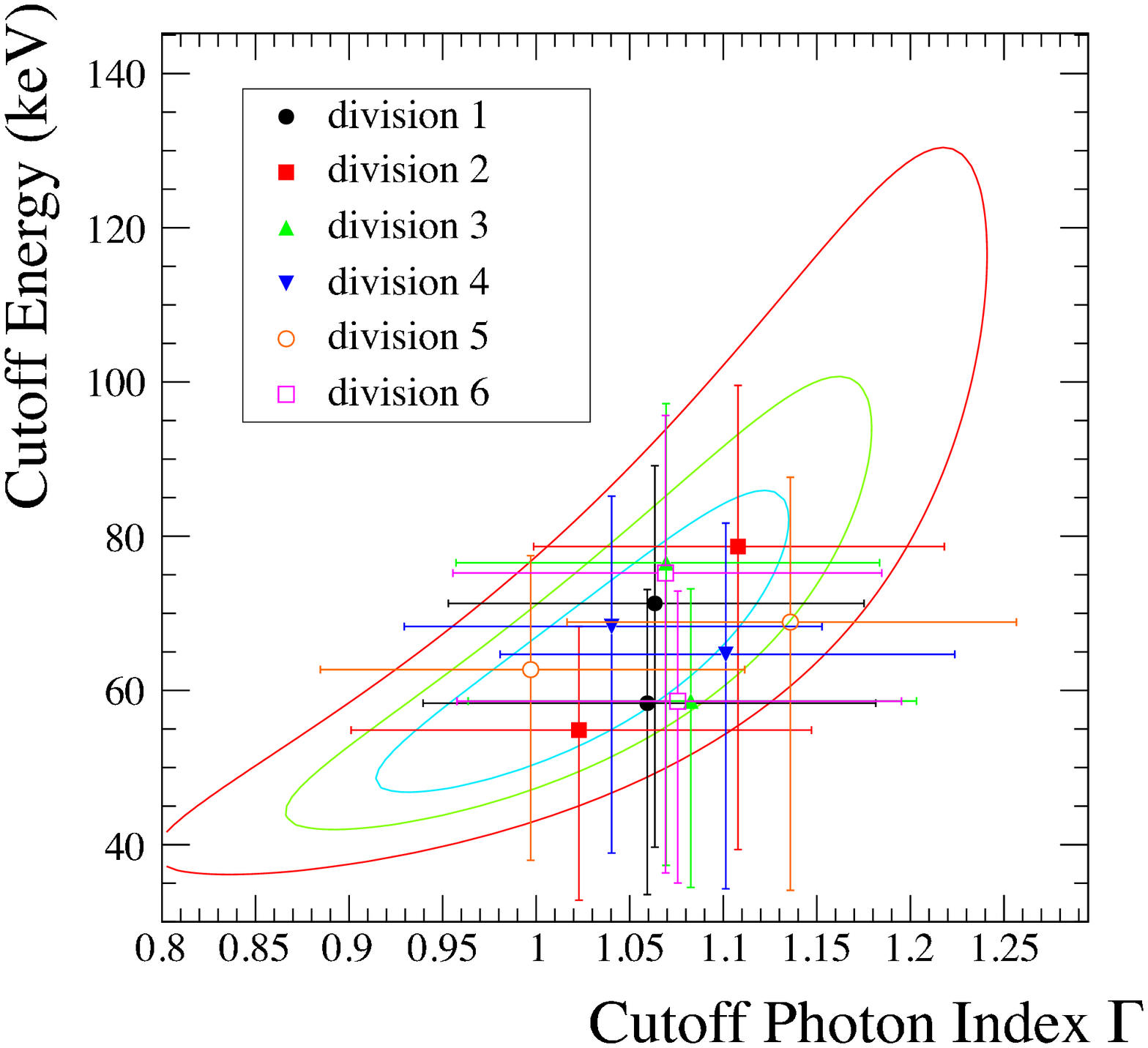}
 \caption{
 \red{
 Confidence contours at the 68\% (light blue), 90\% (green), and 99\% (red) confidence levels 
 	of the CutPL (panel a) 
	fit to the stacked weak burst spectrum.
Individual data points show the parameters,
	with $\pm1 \sigma$ errors, 
	when the 13 short bursts are divided into two groups in six different ways.
}
}
 \label{fig:divide2groups.eps}
\end{figure}

As summarized in Table~\ref{tab: list of spectral fittings of cummulative},	
	a PL model with the fixed $N_{\rm H}$ failed to give an acceptable fit 
	($\chi^2_{\nu}\sim 2.1$; Figure~\ref{fig:accumulated}b),
	while 
	a PL with free $N_{\rm H}$ was more successful ($\chi^2_{\nu}\sim1.3$),
	yielding $\Gamma = 1.57\pm0.04$ and $N_{\rm H}=5.4^{+0.8}_{-0.5}\times 10^{22}$ cm$^{-2}$
	(Figure~\ref{fig:accumulated}c).	
In order to further improve the fit especially in higher energy range,
	we again tried the CutPL and 2BB fits 
	with the same column density fixed at $N_{\rm H}=3.2\times 10^{22}$ cm$^{-2}$.
As summarized in Table \ref{tab: list of spectral fittings of cummulative}
	and shown in Figure~\ref{fig:accumulated}d, 
	the CutPL model gave an acceptable fit ($\chi^2_{\nu}\sim0.8$)
	with $\Gamma = 1.03\pm0.07$
	and 
	$E_{\rm cut}=62.9^{+14.5}_{-10.8}$ keV,
	implying a mild spectral curvature. 
\red{Since $\Gamma$ and $E_{\rm cut}$ couple with each other,
	we show in Fig.~\ref{fig:divide2groups.eps} the fit confidence contours 
	on the $\Gamma$ vs. $E_{\rm cut}$ plane.
In contrast to the successful CutPL model,}
	the more convex 2BB model,
	which was successful on the brightest three bursts 
	(\S\ref{subsection:Characteristics of detected bursts}),
	became much less successful ($\chi^2_{\nu}\sim1.6$; Figure~\ref{fig:accumulated}e).
Thus,
	the weaker bursts are considered to have a flatter 
	10--70 keV HXD-PIN spectrum
	than the brightest bursts,
\red{	particularly Burst 7 and 16. }
	
To make the above spectral difference clearer,
	we added this stacked weak bursts to Figure~\ref{fig:compare_color}a  in $\nu F_{\nu}$ form,
	where we employed the CutPL model for deconvolution in the same way as Burst-7.
The weaker bursts are
	by $\sim$2 orders of magnitude brighter than the persistent emission,
	and 
	by $\sim$1 order of magnitude fainter than Burst-7.
As visualized by this plot,
	the cumulative burst shows a hard X-ray spectrum
	which is less curved than that of Burst 7
	and is similar to that of the persistent X-rays.
In fact,
	the value of $\Gamma_{\rm bst} = 1.57\pm 0.04$, 
	obtained above by the PL fit with free $N_{\rm H}$,
	is consistent with $\Gamma_{\rm per}=1.54^{+0.03}_{-0.04}$ 
	of the persistent hard component 
	(Table~\ref{tab: list of spectral fittings of cummulative}).
Although the CutPL model gave a much harder photon index, $\Gamma_{\rm bst} = 1.03\pm 0.07$,
	the slope is still similar,
	if considering the effect of cutoff around $\sim$63 keV.

In order to more directly compare these $\nu F_{\nu}$ spectra,
	we divided the stacked weak-burst spectrum 
	by that of the persistent emission.
As shown in Figure~\ref{fig:compare_color}b,
	the resultant ratios stay constant at $\sim$170
	over the $\sim$8--200 keV energy range.
The remaining difference in the ratio below $\sim$8 keV in Figure~\ref{fig:compare_color}b 
	is thought to originate mainly from the presence of the soft BB component.
We hence re-calculated the same ratio in Figure~\ref{fig:compare_color}c
	after eliminating the soft BB component from the persistent spectrum.
Thus, the ratio became much flatter even in energies below $\sim$10 keV,
	suggesting that the weak-burst spectrum 
	is similar in shape to the hard X-ray component in the persistent emission.

\red{
Finally, 
	in order to examine our  weak-burst sample for its homogeneity, 
	we subdivided the 13 bursts rather randomly into two subsets, 
	under a constraint that they should be comparable, 
	to within 10\% in the summed signal photon counts (XIS0, PIN, GSO summed). 
From the two subsets,
	we then produced two stacked spectra instead of one,
	and analyzed them in the same manner as before.
The pair of CutPL parameters obtained in this way are indicated in Figure~\ref{fig:divide2groups.eps}
	by a pair  of data ponits with the same symbol.
By trying six different partitions, 
	we confirmed that the two subsets always give,
	within errors,
	consistent model parameters,
	which in turn are also consistent with those derived from the entire sample (indicated by contours).
Therefore, the present sample can be regarded, within errors, as homogeneous.	
}

\subsection{Fitting by thermal/non-thermal components}
\label{subsection: Possible thermal/non-thermal model}
Let us further 
	examine the suggestion of Figure~\ref{fig:compare_color} that 
	the burst emission has the same spectral components as the persistent X-ray emission.
\red{Especially the good similarity above $\sim$8 keV points to
	possible presence of the hard PL component in the stacked weak-burst spectrum,
	like in the persistent emission spectra.}
The short bursts from the transient magnetar SGR~0501+4516
	were already examined for such a possibility
	with an affirmative answer \citep{Nakagawa2011PASJ}.
	
\red{In order to assess the above consideration,
	we re-analyzed the stacked weak-burst spectrum,
	by adding, to the 2BB model, a hard PL
	with its slope fixed at $\Gamma_{\rm per} =1.54$ as specified by the persistent emission (Paper I).}
The inclusion of the hard PL has made 
	the 2BB fit acceptable,
	as summarized 
	in Figure~\ref{fig:accumulated}f  and Table~\ref{tab: list of spectral fittings of cummulative}.
The additional PL contributes to the total 1--300 keV flux by 52\%,
	and is statistically significant,
	because 	
	the achieved fit improvement, $\triangle \chi^2=-27.1$,
	implies an F-test chance probability of of $2.4\times 10^{-7}$.
\red{Even though an acceptable fit with a comparable goodness
	had already been obtained with the CutPL model,
	we consider the 2BB+PL modeling more appropriate,
	for the following two reasons.
One is that this modeling has successfully explained the spectra 
	of weak short bursts from SGR~0501+4516 \citep{Nakagawa2011PASJ}.	
The other is that the 2BB model has been regarded as a standard representation of short-burst spectra
	of a fair number of magnetars (\citealt{Nakagawa2007PASJ}  and references therein),
	so that our final model should also be based on it, 
	rather than the CutPL model which in the present case is quite conventional. 	}

\begin{figure}
\begin{center}
\includegraphics[width=85mm]{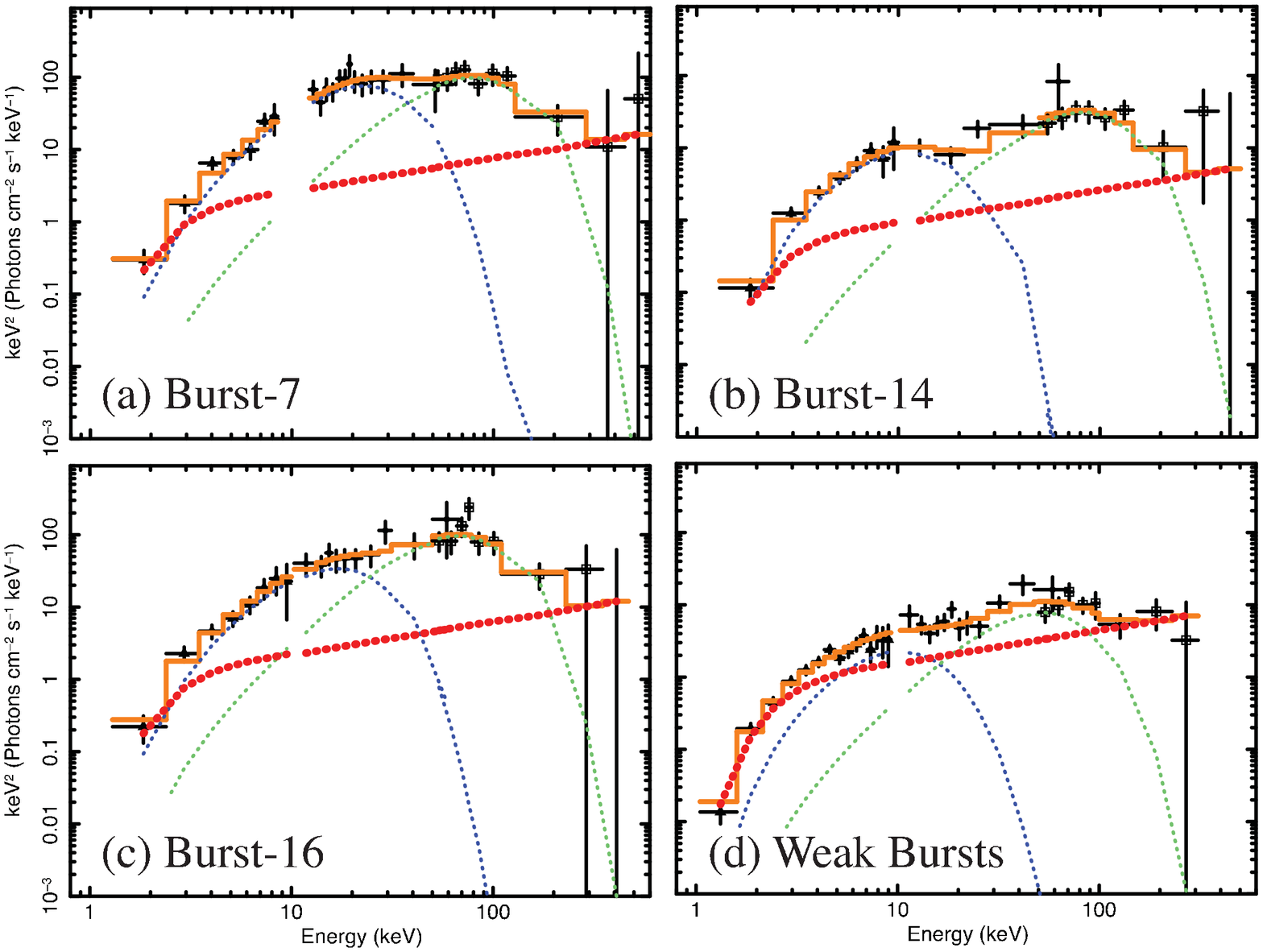}
 \caption{
An $\nu F_{\nu}$ presentation of 
	the 2BB plus PL (with $\Gamma =1.54$ fixed)
	fits to the spectra of 
	Burst-7 (panel a), Burst-14 (panel b), Burst-16 (panel c), 
	and the stacked weak short bursts (panel d).
The two BB components are represented in blue and green,
	while the hard PL in red.
 }
  \label{fig:2bb_pl.eps}
\end{center}  
\end{figure}

\red{
Now that evidence for the PL component has been obtained in the stacked weak-burst spectrum,
	it would be a natural step to examine the brighter ones,
	Burst-7, 14, and 16, for the same possibility,
	even though their spectra were already represented successfully by the 2BB model.
This is because the distinction between the brighter and fainter bursts in the present observation 
	is purely technical, without particular reason to consider that the two groups have distinct nature.
	Then, to the 2BB model describing the 3 bright bursts,}
	we added the hard PL with $\Gamma_{\rm bst}$ fixed at the same value.
This improved the fit to 
	Burst-7, 14, and 16,
	by $\triangle \chi=-4.3$, $-2.0$, and $-2.8$, 
	giving F-test probabilities of 
	0.60\%,
	7.4\%,
	and 
	2.4\%, respectively.
Thus, 	
	the hard PL component is likely to contribute at least to Burst-7.
The best fit spectra are shown in Figure~\ref{fig:2bb_pl.eps} in $\nu F_{\nu}$ forms,
	with the derived parameters in Tabel~\ref{tab: list of spectral fittings}.
The BB temperatures did not change within errors from those 
	obtained in the previous pure 2BB modeling 
	(\S\ref{subsection:Characteristics of detected bursts}),
	while  
	the hard PL component was found to contribute
	to the total 1--300 keV fluxes of 
	Burst-7, 14, and 16,
	by 10\%, 14\%, and 19\% respectively.

\section{DISCUSSION}
\label{discussion}

\subsection{Spectral comparison among different emissions}
\label{subsection: Spectral comparison among different emissions}

During the 33.5 ks of effective exposure with the HXD 
	onto 1E~1547.0$-$5408 in its 2009 January activity,
	we detected 18 short bursts with their fluence $S$
	above $S_{\rm min}\sim 2\times 10^{-9}$ erg cm$^{-2}$,
	and analyzed 16 of them. 
Even the strongest three among the 16 bursts, 
	with the 10--70 keV fluence of $S=4$--$8\times 10^{-8}$ erg cm$^{-2}$
	(or $\sim 1 \times 10^{-7}$ erg cm$^{-2}$ in 1--300 keV),
	are still weaker than most of previously reported bursts 
	from magnetars in general
	which typically have $S \ga 10^{-7}$ erg cm$^{-2}$.
For example, 
	the burst forest of 1E~1547.0-5408 observed on 2009 January 22 
	with {\it Swift}, {\it INTEGRAL}, {\it Suzaku}, and {\it Fermi}
	was composed of events
	with $S\ga 2\times 10^{-6}$ erg s$^{-1}$ up to $2.5	\times 10^{-4}$ erg cm$^{-2}$ \citep{Mereghetti2009ApJ}.
Therefore, the present {\it Suzaku} results,
	together with those by \cite{Nakagawa2011PASJ} on SGR~0501+4516,
	provide valuable information on the wide-band spectra of weak short bursts.
For reference,
	the present sample defines a burst frequency of 
	$5.4\times 10^{-4}$ burst s$^{-1}$ in the range of $S \ga S_{\rm min}$.
	 
So far, 
	wide-band spectra of many energetic ($S> 10^{-7}$ erg s$^{-1}$) short bursts 
	from magnetars 
	have been explained successfully by the 2BB model:
	these include numerous bursts from SGR~1806$-$20 and SGR~1900$+$14 \citep{Nakagawa2007PASJ, Israel2008ApJ},
	as well as the strongest {\it Suzaku} burst from SGR~0501+4516
	\citep{Enoto2009ApJ,Nakagawa2011PASJ}.
At the same time,
	the 2BB model can generally reproduce the soft X-ray component,
	which is ubiquitously seen at energies of $\la 10$ keV of wide-band persistent spectra from magnetars
	\citep{Mereghetti2008A&ARv}
	including 1E~1547.0$-$5408 itself (Paper I).
These 2BB fits to the two different forms of magnetar emission reveal an interesting 
	common scaling as $T_{\rm Low}/T_{\rm High} \sim 0.4$ \citep{Nakagawa2009PASJ},
	where $T_{\rm Low}$ and $T_{\rm High}$ denote 
	the lower and higher 2BB temperatures, respectively.
These results reinforce the possibility 
	that the persistent emission is composed 
	of numerous micro short bursts (\S\ref{Introduction}; \citealt{1996ApJ...473..322T, Lyutikov2003MNRAS, Nakagawa2007PhD}).
Since the 2BB modeling is thus considered to have some physical meanings 
	beyond a mere convention,
	below we construct our discussion based on the 2BB modeling,
	rather than those with the CutPL model which is rather empirical.

The three brightest bursts in our present sample,
	\red{particularly  the brighter two (Burst 7 and 16),}
among our sample exhibit upward convex spectral shapes,
	which have been successfully described by the 2BB model
	over an extremely wide energy band of $\sim 1$ to $\sim 300$ keV
	(\S\ref{subsection:Characteristics of detected bursts}).
Importantly, the values of $T_{\rm Low}$ and $T_{\rm High}$ 
	derived from these bursts (possibly except for Burst-14) obey the above scaling relation.
In contrast,
	the stacked weaker bursts have a less convex shape (Figure \ref{fig:compare_color}),
	and 	did not accept the simple 2BB modeling.
This is understandable if we presume  
	that the spectra of weaker bursts are contributed by 
	the hard component with a power-law photon index $\Gamma \sim 1$,
	which is generally observed in the persistent emission of magnetars 	
	\citep{Kuiper2006ApJ,Enoto2010ApJ}.
The weak bursts from SGR~0501+4516,
	when stacked together, actually exhibited a hard-tail feature \citep{Nakagawa2011PASJ}.
\red{In fact,
	Figure \ref{fig:compare_color} b and c  provide supporting evidence for this possibility,
	where the ratios become flatter above $\sim$8 keV.}
We have accordingly added a hard 	PL component to the 2BB model in \S\ref{subsection: Possible thermal/non-thermal model},
	fixing $\Gamma$ at the same value as obtained from the persistent emission,
	and 
	obtained a fully acceptable spectral fit.
Thus, the hard component 	
	is significantly present in the stacked weak-burst spectrum,
	as long as the 2BB model is chosen as the start point of analysis.
\red{As a further justification of this 2BB+PL modeling, 
	the stacked spectrum yield $T_{\rm High}/T_{\rm Low}=4.9\pm1.1$ 
	(from Table~\ref{tab: list of spectral fittings of cummulative})
	in agreement with the ''canonical" ratio,
	with it was $7.3\pm0.5$ before properly including the PL component.
}
As argued by \cite{Nakagawa2011PASJ},
	these strengthen the similarity between the burst and persistent emissions.
The report by \cite{Israel2008ApJ},
	that weaker bursts tend to show harder spectra in terms of the 2BB analysis,
	may be understood if considering possible contributions of 
	the same PL component to their weaker bursts.	

\begin{figure}
\includegraphics[width=80mm]{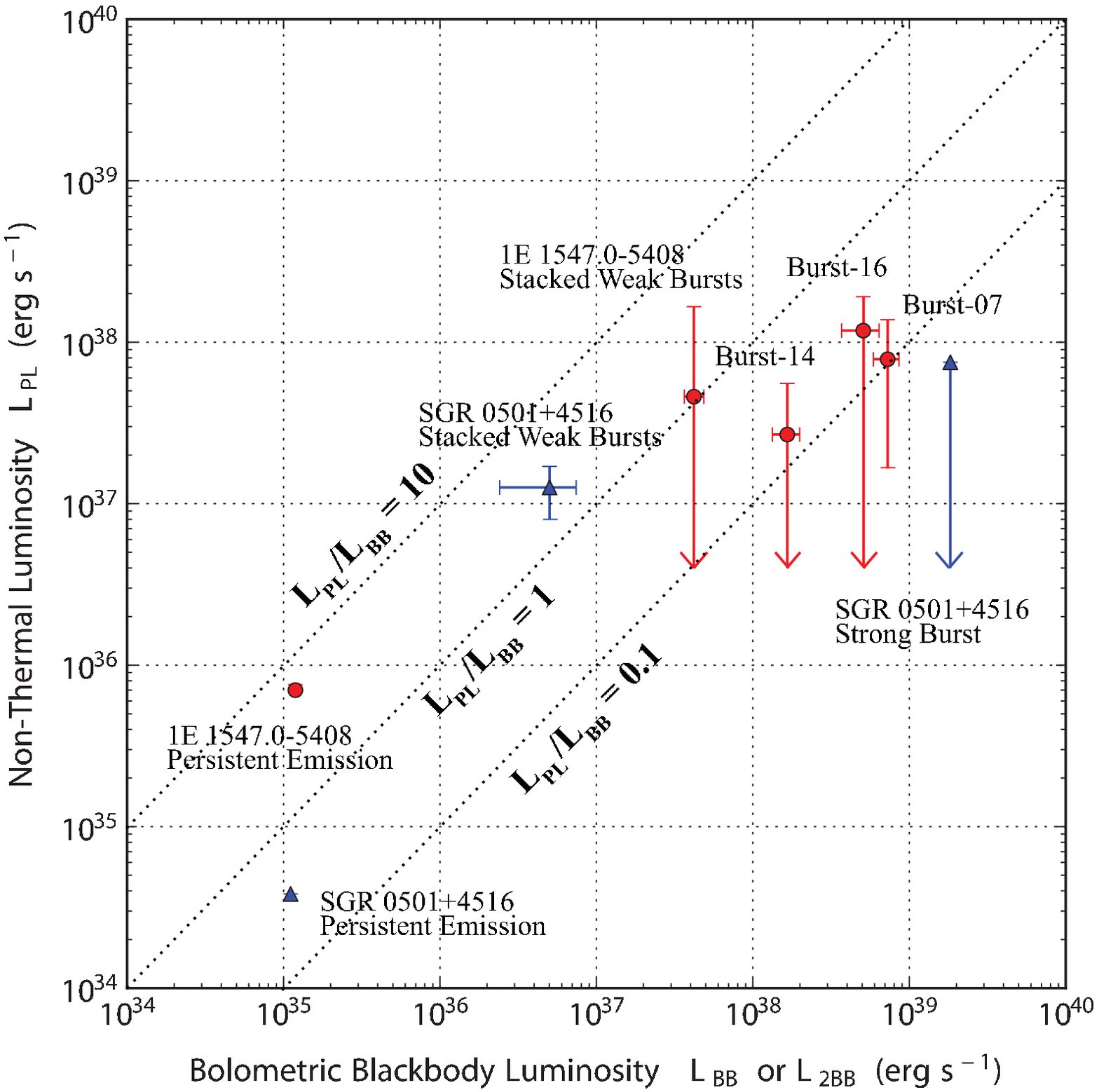}
 \caption{ Comparison of the 1--300 keV PL luminosity to 
        the bolometric BB luminosity.
The results on SGR~0501+4516 were taken from Figure~8 in \citet{Nakagawa2011PASJ}.
In this plot, errors are shown at the 90\% C.L. level. }
  \label{fig:thermal_vs_nonthermal.eps}
\end{figure}

The above results have lead us to a natural question about brighter short bursts,
	of which the spectra generally lack the PL component:
	is this an intrinsic effect due, e.g., to some changes in the emission mechanism,	
	or an artifact caused by their higher 2BB temperatures which mask any PL emission?
With this in mind,
	we examined (\S\ref{subsection: Possible thermal/non-thermal model})	
	the spectra of the three brighter {\it Suzaku} bursts,
	for possible contribution from such a hard component. 
Then, 
	the presence of the PL component (on top of the 2BB model)
	was suggested 
	in Burst-7 
	at a relatively high confidence ($>$97\%) level.
The results are presented in Figure \ref{fig:thermal_vs_nonthermal.eps}
	on the thermal (2BB) vs non-thermal (PL) luminosity plane,
	in comparison with the weak-burst and persistent spectra.
There,	
	assuming a source distance of 4 kpc \citep{Tiengo2010ApJ}, 
	the persistent X-ray emission (Paper I)
	is represented by 
	a bolometric blackbody luminosity of 
	$L_{\rm BB}=1.2\times 10^{35}$ erg s$^{-1}$	
	and absorption-corrected 1--300 keV PL luminosity of 
	$L_{\rm PL}=6.9\times 10^{35}$ erg s$^{-1}$,
	which sum up to give the total luminosity of 
	$L_{\rm per}=8.1\times 10^{35}$ erg s$^{-1}$.
We also quote the results on SGR~0501+4516,
	taken from \cite{Nakagawa2011PASJ}.
Although it is still difficult at present to distinguish the two alternatives
	the PL luminosity, \red{if it is universally present in short bursts}, 
	appears to saturate at $\sim 10^{38}$ erg s$^{-1}$,
	preferring the former possibility. 

 On large scales, 
 	Figure~\ref{fig:thermal_vs_nonthermal.eps} reveals a clear positive
	correlation between luminosities of the two emission components.
Therefore, 
	the soft (thermal) and hard (PL) components
	are considered to approximately keep their luminosity ratio over a very 
	broad scale in flux or fluence. 
Similarly, \cite{Enoto2010ApJ} showed
	that the luminosity ratio between these two emission components,
	comprising the persistent emission, is not much different between
	magnetars in activity and those in quiescence. 
These facts give another
	support to the microburst conjecture.

\subsection{LogN-logS relation of short burst}
\label{LogN-logS relation of short burst}
In order to examine the possibility 
	that the persistent X-ray emission is composed of micro burst events,
	it is inevitable to evaluate X-ray fluxes accumulated over resolved and unresolved bursts events.
Since the present short bursts, with $S\ga S_{\rm min}\sim 2\times 10^{-9}$ erg cm$^{-2}$,
	were detected at a low burst frequency, $5\times 10^{-4}$ s$^{-1}$,
	their accumulated flux is still at a level of $\sim 10^{-12}$ erg s$^{-1}$ cm$^{-2}$.
This is by $\sim$2 orders of magnitude lower than the observed persistent X-ray flux.
Therefore, 	
	much larger contributions of smaller unresolved short bursts
	are required to explain the persistent X-ray flux.

For a quantitative estimate,
	let $N(>S)$ denote the occurrence frequency of those bursts of which the fluence is $>S$.
Observationally, $N$ is thought to be described by a single power-law as 
	$N(>$$S) \propto S^{-\alpha}$,
	like in solar flares and earthquakes.
Sometimes known as  the Gutenberg-Richter law,
	this relation is considered to represent self-organized criticality.
Previous observations of SGR~1806$-$20 gave a range of $\alpha=0.7$--$1.1$; 
	e.g.,
	$\alpha \sim 0.9$ by the Konus-{\it Wind} \citep{Aptekar2001ApJS},
	$\alpha = 0.76 \pm 0.17$ by the BATSE, 
	$\alpha = 0.67\pm 0.15$ by ICE \citep{Gogus2000ApJ},
	$\alpha = 0.91\pm0.09$ by {\it INTEGRAL} \citep{Gotz2006A&A},
	and $\alpha=1.1\pm 0.6$ by {\it HETE}-2 \citep{Nakagawa2007PASJ}.
From the present target, 1E~1547.0$-$5408,	
	the slope is reported to be
	$\alpha=0.75\pm0.06$	 
	and 
	$\alpha=0.6\pm 0.1$ 	
	through 
	{\it INTEGRAL}/ACS \citep{Mereghetti2009ApJ} and
	{\it Swift}/XRT \citep{2011arXiv1106.5445S} 
	observations, respectively.

\begin{figure}
\begin{center}
\includegraphics[width=85mm]{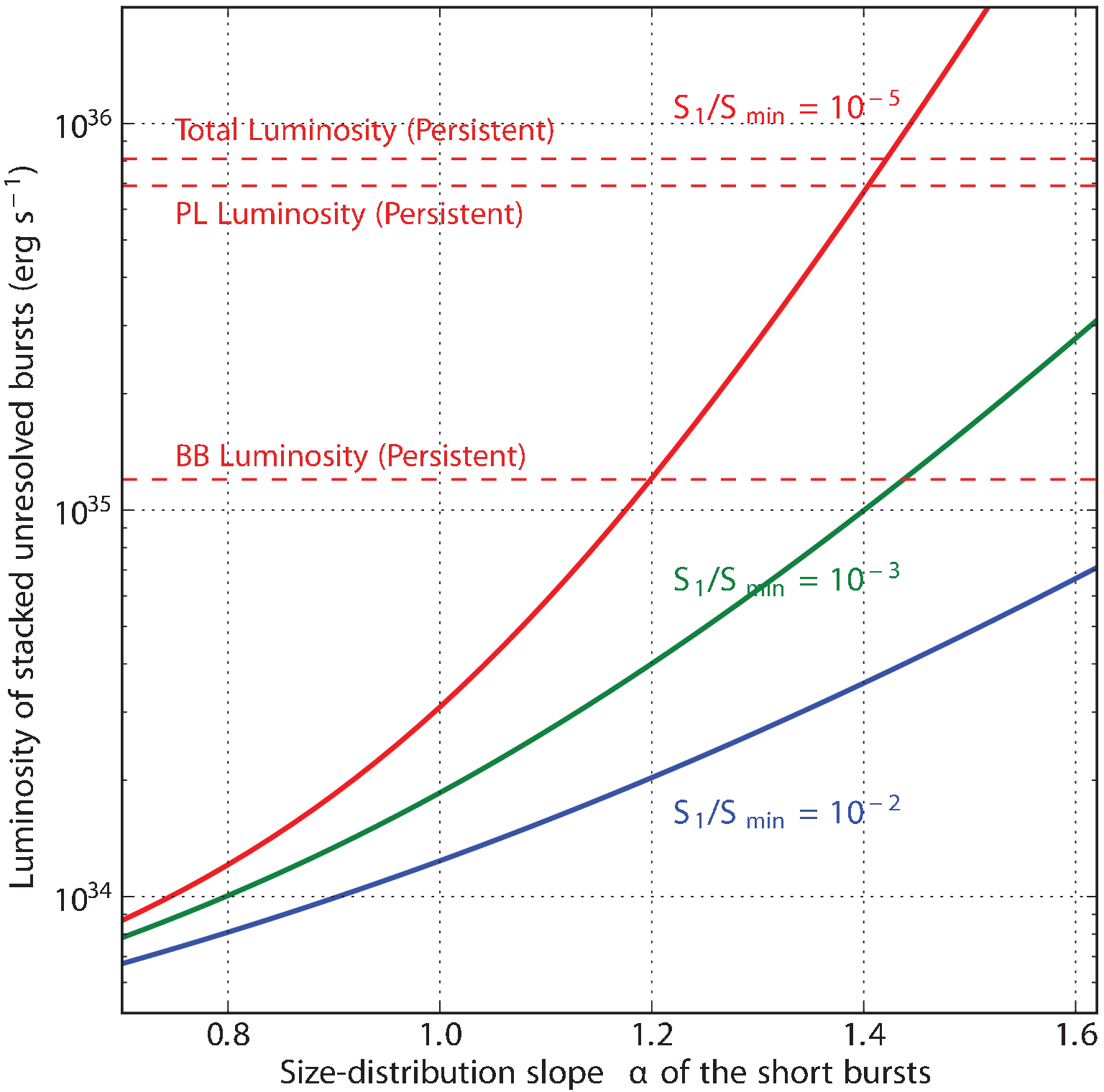}
 \caption{
Expected luminosity of integrated unresolved weaker bursts,
	as a function of the size-distribution slope $\alpha$.
Blue, green, and red lines are the calculations 
	with the lower bounds of the integration
	at $S_1/S_{\rm min}=10^{-2}$, $10^{-3}$, and $10^{-5}$, respectively.
Three horizontal lines represents the persistent X-ray luminosity of 1E 1547.0-5408
	for the total (top), PL (middle),  and BB (bottom) components.
 }
  \label{fig:luminosity_alpha_GL.eps}
\end{center}  
\end{figure}
	
The burst number density at a particular fluence $S$
	is written as $f_{\rm bst}(S) \propto -dN(S)/dS  \equiv (f_0/S_0) (S/S_0)^{-\alpha-1}$,
	where $S_0$  is a typical fluence and 
	 $f_0$ (burst s$^{-1}$) is a normalization constant.
Thus, the luminosity integrated over bursts between $S_1$ and $S_2$,
	 at a distance $d$ (kpc), is given as 
\begin{eqnarray*}
L_X &=& \int_{S_1}^{S_2} 4\pi d^2 S f_{\rm bst}(S) d(S) \\
	&=& 4\pi d^2S_0f_0 \int_{s_1}^{s_2}  s^{-\alpha}ds \\	
	&=& 4\pi d^2S_0f_0 (s_2^{1-\alpha}-s_1^{1-\alpha})/(1-\alpha) ,
\end{eqnarray*}
with $s\equiv S/S_0$, $s_1\equiv S_1/S_0$,  and $s_2\equiv S_2/S_0$.
Here, let us assume 
	$S_0=S_{\rm min}\sim 2\times 10^{-9}$ erg cm$^{-2}$ (the present detection limit),
	and $s_2=S_2/S_{\rm min}=1$,
	at $d=$4 kpc.
The observed frequency $N(S_{\rm min})\sim 5.4\times 10^{-4}$ burst s$^{-1}$
	gives 
	 $f_0  \sim 7\times 10^{-4}$ burst s$^{-1}$.
In order to explain the persistent luminosity in this way,
	we clearly need $\alpha>1.0$,
	because $S\times N\propto S^{1-\alpha}$ should increase
	towards smaller values of $S$.
More quantitatively,
	Figure~\ref{fig:luminosity_alpha_GL.eps}
	shows the calculated value of $L_{\rm X}$ as a function of  $\alpha$ 
	for these different values of $s_1$.

Comparing the three brightest bursts and the stacked weaker ones
	in Table~\ref{tab: list of spectral fittings}
	and
	Table~\ref{tab: list of spectral fittings of cummulative},
	the blackbody radii 
	of neither the lower- or higher-temperature components
	are strongly dependent on their fluences $S$.
Thus,
	as a first-order approximation,
	let us assume $S\propto kT^4$ after the Stefan-Boltzmann law.
Since 
	the weak bursts with $S_{\rm min}$
	and 
	persistent emission 
	give
	$kT_{\rm High}\sim 12$ keV 
	and 
	$kT \sim 0.65$ keV,
	respectively,
	a typical unresolved weakest burst 
	is considered to have a fluence 
	$s_1 = S_1 / S_{\rm min} \sim (0.65 {\rm \ keV}/12 {\rm \ keV})^4 \sim 10^{-5}$.
Combining with Figure~\ref{fig:luminosity_alpha_GL.eps},
	this implies that the slope $\alpha$ should be $\sim$1.2 and $\sim$1.4
	to supply the luminosity of the persistent BB and PL luminosities, respectively.

The required value of  $\alpha=1.2$--1.4
	is 
	larger than has been measured.
Since the observations of bursts
	were performed mainly during more burst-active states than in the {\it Suzaku} observation,
	a possible interpretation is variations of $\alpha$ depending on the magnetar activity.
For example, 
	the solar flare is known to change its size-distribution slope $\alpha$,
	from $\sim$0.8 at activity peaks
	to $\sim$1.2 in more quiescent states
	\citep{Bromund1995ApJ...455..733B}.
This is also pointed out in some magnetars (see also SGR~1806$-$20 and SGR~1900+14 in \citealt{Nakagawa2007PASJ}, related with SGR~0501+4516 behavior in \citealt{Gogus2010ApJ}).
If $\alpha$ reaches 1.2--1.4
	in less burst-active states,
	the persistent emission can be explained by the superposition of micro bursts.
Otherwise,
	an alternative dissipation process that is independent of the short-burst production 
	is inferred to be also contributing to the persistent emission.
Such processes include, 
	e.g., 
	internal crustal heating in magnetars \citep{2008ApJ...673L.167A}.

\subsection{Energy source for the emissions}
\label{Energy source for the emissions}

Finally,
	let us compare energetics 
	of the bursts
	and 
	persistent emission 
	with the magnetic energies of magnetars. 
Assuming 4 kpc as before, 
	the 2BB model and the Stefan-Boltzmann law
	measure the total radiated energies of 
	$2.1\times 10^{38}$ erg,
	$1.6\times 10^{38}$ erg,
	and 
	$1.6\times 10^{38}$ erg,
	from 
	 Burst-7,
	 Burst-14,
	 and Burst-16, respectively.
Further considering the $T_{90}$ values,	 
	these yield bolometric luminosities of 
	$7.8\times 10^{38}$ erg s$^{-1}$,
	$1.9\times 10^{38}$ erg s$^{-1}$,
	and
	$5.9\times 10^{38}$ erg s$^{-1}$,	
	respectively.
In the weaker bursts,
	an average energy emitted per  bust 
	becomes
	$1.2\times 10^{37}$ erg, 
	and 
	the bolometric luminosity is
	$4.2\times 10^{37}$ erg s$^{-1}$.


In comparison with the above estimates, 
	the magnetic energy	stored within a size of $R_{\rm mag}$ becomes
\begin{eqnarray*}
E_{\rm bst}&=& \frac{B^2}{8\pi}R_{\rm mag}^3 {\rm \ erg} \\
	&=&2 \times 10^{45} \left(\frac{B}{B_{\rm surf}}\right)^2 
	\left(\frac{R_{\rm mag}}{R_{\rm NS}}\right)^3 {\rm \ erg},
\end{eqnarray*}	
when normalizing to the dipole magnetic field, $B_{\rm surf}=2.2\times 10^{14}$ G \citep{Camilo2007ApJ},
	and 
	employing 
	the canonical neutron start radius $R_{\rm NS}=10$ km.
If the individual burst emission is powered by dissipation of 
	the magnetic energy in this region (e.g., via reconnection; \citealt{Lyutikov 2006MNRAS}),
	a typical size of $R_{\rm mag} \ga10^{-2.3}R_{\rm NS}$ is needed.
Blackbody radii observed in our sample (the 2BB model),
	$R_{\rm Low}$$\sim$1 km = $0.1R_{\rm NS}$
	and 
	$R_{\rm High}$$\sim$0.1 km = $10^{-2}R_{\rm NS}$,	
	are larger than the required $R_{\rm mag}$.
This is considered reasonable,
	because the released magnetic energy would in any case diffuse out,
	and would make $R_{\rm BB}>R_{\rm mag}$.

If we assume $s_1\sim 10^{-5}$ and $\alpha \sim 1.4$
	like in the previous subsection,
	the burst frequency at $S_1$ is estimated to be $f\sim 10^{3}$ Hz.
Therefore, a minimum size of $R_{\rm mag}/R_{\rm NS} \sim 10^{-4}$ is enough to 
	supply the observed luminosity $L_{\rm x}\sim 6\times 10^{35}$ erg s$^{-1}$,
	if the released magnetic energy is evaluated as 
	$L_{\rm mag}  \sim 2 \times 10^{45} (B/B_{\rm surf})^2 (R_{\rm mag}/R_{\rm NS})^3 f$ erg s$^{-1}$.
Since the persistent emission exhibits 
	$R_{\rm BB} \sim 5$ km (Paper I),
	the condition of $R_{\rm BB}>R_{\rm mag}$ is retained.

\section{CONCLUSION}
\label{conclusion}

We studied short bursts from 1E~1547.0$-$5408 (also known as SGR~J1550$-$5418, PSR~J1550$-$5418) 
	detected during the {\it Suzaku} pointing observation 
	on 2009 January 28--29 (UT)
	in its burst-active state.
This period was 7 days after the burst forests on January 22.
Combined with the previous study of its persistent X-ray emission (Paper I),
	we have obtained the following results.	

\begin{enumerate}
\item 
Using a $\triangle t$ distribution of the HXD-PIN data,
	we identified 18 short burst events 
	with its risk probability  below $\sim 1.7\times 10^{-6}$.
Sixteen of them,  free from the data loss,
	were used for the spectral analyses.
	
\item 	
Three brightest bursts have their fluences of $\sim 4$--$8\times 10^{-8}$ erg cm$^{-2}$
	in the 10--70 keV energy range.
Their individual spectra 
	were fitted successfully by
	the 2BB model 
	(and some other models)
	over the 0.5--$\sim$400 keV band.

\item 
Remaining 13 bursts 
	define one of the weakest samples ever measured,
	with their fluences covering the range of $2\times 10^{-9}$--$2\times 10^{-8}$ erg cm$^{-2}$.
Their stacked spectrum shows 
	similarity to the persistent emission
	in its slope above $\sim$8 keV.
	\red{This spectral similarity extends to even towards lower energies after eliminating the BB component from the persistent spectrum.}

\item
The stacked spectrum cannot be represented by a 2BB model, 
	while an additional hard PL made the fit acceptable,
	even fixing the slope at the value of the persistent one.
The luminosity of the hard PL shows a sign to saturate at $\sim$10$^{38}$ erg s$^{-1}$,
	when compared to the thermal luminosity.

\item	
We evaluated the available energy supplied from the unresolved short bursts.
The persistent emission from 1E~1547.0$-$5408 
	can be explained if the slope $\alpha$ of the cumulative distribution of the short bursts
	is rather steep as $\alpha=1.2$--$1.4$.
\end{enumerate}

We thank members of the {\it Suzaku} magnetar Key Project
	and the {\it Suzaku} operation teams for the successful ToO.
TE is supported by the JSPS Postdoctoral Fellowships for Research Abroad. 




\begin{table*}
\begin{flushleft}
\caption{List of detected short burst events.}
\label{tab: list of burst}
\begin{tabular}{@{} rlrrccclllc}
\hline \hline
ID & Time at Burst Peak & T90 & $N_{\rm bst}^{1}$ & $N_{\rm T90}^{2}$ & 1-sec Rate$^2$ & 
Model &   Photon  & Flux$^3$ & Fluence$^3$ & $\chi_{\nu}^2$  \\
& (UTC) & (ms) & &  &  & & Index & & &(p-value) \\
\hline
1 & 2009-01-28T21:48:24.8 & 1851.4 & 9 & (72, 26, 29) & (37, 15, 25) & PL & $1.65_{-0.13}^{+0.16}$ & $0.5_{-0.1}^{+0.1}$ & $9.0_{-2.4}^{+1.1}$ & 1.09(0.36) \\
2 & 2009-01-29T02:36:02.2 & 46.9 & 9 & (16, 9, 4) & (26, 10, 25)  & PL & $1.34_{-0.12}^{+0.13}$ & $6.0_{-1.6}^{+1.0}$ & $2.8_{-0.7}^{+0.5}$ & 0.81(0.60) \\
3 & 2009-01-29T03:06:21.3 & 148.4 & 5 & (9, 4, 7) & (20 , 6, 14) & PL & $1.07_{-0.25}^{+0.29}$ & $1.3_{-0.7}^{+0.3}$ & $1.9_{-1.1}^{+0.5}$ & 0.75(0.69)\\
4 & 2009-01-29T08:09:06.6 & 226.5 & 4 & (17, 8, 10) & (21, 10, 21) &  PL  & $1.24_{-0.12}^{+0.12}$ & $1.5_{-0.4}^{+0.5}$ & $3.4_{-0.9}^{+1.2}$& 1.19(0.31) \\
5 & 2009-01-29T08:53:23.4 & 78.1 & 5 & (8, 5, 4) & (12, 9, 14)  &  PL & $1.18_{-0.17}^{+0.19}$ & $2.4_{-1.1}^{+0.5}$ & $1.9_{-0.8}^{+0.4}$ & 0.44(0.82)  \\
6 & 2009-01-29T09:17:10.4 & 656.2 & 256 & (2029, 219, 59) & (2107, 282, 93) &  PL  & -- & -- & --  & -- \\ 
7 & 2009-01-29T11:19:27.4 & 265.6 & 139 & (127, 131, 84) & (155, 143, 101) & CutPL & $0.12_{-0.14}^{+0.13}$ & $28.1_{-3.4}^{+0.7}$ & $74.6_{-9.1}^{+1.9}$ & 0.82(0.74)\\
8 & 2009-01-29T12:16:56.4 & 390.6 & 19 & (44, 23, 15) & (56, 28, 26) & PL  & $1.49_{-0.07}^{+0.08}$ & $1.8_{-0.4}^{+0.4}$ & $7.1_{-1.5}^{+1.4}$& 1.46(0.15) \\
9 & 2009-01-29T12:17:36.5 & 54.7 & 11 & (19, 11, 10) & (34, 16, 32)  &  PL & $0.91_{-0.13}^{+0.13}$ & $15.0_{-4.3}^{+1.9}$ & $8.2_{-2.3}^{+1.0}$& 0.28(0.95) \\
10 & 2009-01-29T12:58:41.1 & 78.1 & 44 & (--, 38, 29) & (--, 47, 41)  & PL   &  -- & -- & -- & -- \\ 
11 & 2009-01-29T13:41:57.1 & 164.1 & 8 & (21,  8, 11) & (37, 10, 20) & PL   & $1.45_{-0.12}^{+0.14}$ & $2.5_{-0.6}^{+0.5}$ & $4.1_{-0.9}^{+0.8}$& 0.85(0.55)  \\
12 & 2009-01-29T13:43:27.9 & 210.9 & 7 & (7, 9, 13) & (13, 10, 26)  & CutPL &  $-0.38_{-0.55}^{+0.39}$ & $4.0_{-0.7}^{+0.8}$ & $8.4_{-1.4}^{+1.7}$& 0.53(0.78) \\
13 & 2009-01-29T13:46:37.4 & 328.1 & 13 & (97, 20, 19) &  (113, 27, 26) & CutPL &  $1.11_{-0.18}^{+0.17}$ & $4.3_{-1.3}^{+0.2}$ & $14.0_{-4.3}^{+0.7}$ & 0.87(0.56)\\
14 & 2009-01-29T13:50:38.2 & 835.9 & 60 & (156, 62, 82) & (187, 67, 89) & CutPL & $0.83_{-0.09}^{+0.09}$ & $5.5_{-0.8}^{+0.3}$ & $45.9_{-6.4}^{+2.5}$& 1.10(0.34) \\
15 & 2009-01-29T15:40:45.1 & 109.4 & 15 & (35, 14, 14) &  (44, 16, 24) &  PL & $1.48_{-0.08}^{+0.09}$ & $4.8_{-0.9}^{+0.7}$ & $5.3_{-1.0}^{+0.7}$& 1.23(0.26) \\
16 & 2009-01-29T16:51:40.5 & 265.6 & 90 & (97, 85, 75) & (148, 96, 98)  & CutPL & $0.28_{-0.12}^{+0.12}$ & $21.1_{-2.9}^{+0.9}$ & $56.1_{-7.8}^{+2.4}$& 0.54(0.97)\\
17 & 2009-01-29T16:59:58.8 & 109.4 & 6 & (19,  6,  5) & (24, 8, 19) & PL  &  $1.39_{-0.15}^{+0.17}$ & $2.6_{-1.1}^{+0.7}$ & $2.9_{-1.2}^{+0.8}$& 0.78(0.60)\\
18 & 2009-01-29T18:42:12.6 & 54.7 & 5 & (17,  6, 7) & (25 , 8, 23)  & PL  &  $1.38_{-0.16}^{+0.19}$ & $6.5_{-1.7}^{+1.6}$ & $3.6_{-1.0}^{+0.9}$& 0.27(0.97) \\
\hline
\end{tabular}

\medskip
1: An event train of length $N_{\rm bst}$ is defined in \S\ref{subsection:Delta-time distribution}. \\ 
2: $N_{\rm T90}$ and 1-sec rate around the burst are shown in a form of (XIS0, PIN, GSO). The former is a burst photon count during $T_{\rm 90}$ defined in \S\ref{subsection:Characteristics of detected bursts}. The latter is extracted from 1-sec binned light curves in Figure~ \ref{fig:light_curves}. \\
3: Flux is shown in a unit of $10^{-8}$ erg s$^{-1}$ cm$^{-2}$, and fluence is in $10^{-9}$ erg cm$^{-2}$ both in the 10--70 keV band. \\
\end{flushleft}
\end{table*}

\begin{table*}
\begin{flushleft}
\caption{Spectral parameters of the three brightest short bursts. 
}
\label{tab: list of spectral fittings}
\begin{tabular}{@{}lllll}
\hline \hline
Spectral Model & Parameter & BST-7 & BST-14 & BST-16 \\
\hline
& Time (2011-01-29) &
11:19:27.4 & 
13:50:38.2 & 
16:51:40.5
\\
& T90 (sec) & 
0.2656 &
0.8359 & 
0.2656 
\\

& Fluence\footnotemark ($10^{-8}$ erg cm$^{-2}$) &  
$7.46^{+0.48}_{-0.82}$ &
$4.60^{+0.25}_{-0.67}$ &
$5.60^{+0.24}_{-0.72}$
\\ 

\hline
wabs\footnotemark
& $N_{\rm H}$ ($10^{22}$ cm$^{-2}$) 
& (3.2 fix) & (3.2 fix) & (3.2 fix) \\
\hline 
CutPL &

Photon Index $\Gamma$ & 
$0.12^{+0.13}_{-0.14}$ &
$0.83\pm 0.09$ &
$0.28\pm 0.12$ 
\\

& Cutoff Energy $E_{\rm cut}$ (keV) & 
$26.9^{+4.4}_{-3.8}$ &
$65.7^{+16.7}_{-12.3}$ &
$31.9^{+4.8}_{-4.1}$
\\


& PIN Flux\footnotemark ($10^{-7}$ erg s$^{-1}$ cm$^{-2}$) & 
$2.81^{+0.18}_{-0.31}$ & 
$0.55^{+0.03}_{-0.08}$ & 
$2.11^{+0.09}_{-0.27}$
 \\
 
 & Total Flux\footnotemark  ($10^{-7}$ erg s$^{-1}$ cm$^{-2}$) & 
$4.01^{+0.14}_{-0.51}$ &
$1.08^{+0.07}_{-0.18}$ &
$3.21^{+0.11}_{-0.56}$
 \\
 
& fit goodness $\chi^2_{\nu}$ ($\nu$) & 
0.82 (28) &
1.10 (20) &
0.54 (25)
\\

& figure & 
fig.\ref{fig:moderate-intensity_short_burst_spec}(a1,2,5) & 
fig.\ref{fig:moderate-intensity_short_burst_spec}(b1,2,5) & 
fig.\ref{fig:moderate-intensity_short_burst_spec}(c1,2,5)
\\

\hline
2BB 
& $kT_{\rm Low}$ (keV) &
$5.28^{+0.43}_{-0.42}$ &
$2.58_{-0.34}^{+0.47}$ &
$3.97\pm 0.41$
\\


& $R_{\rm Low}$ (km) & 
$1.78^{+0.20}_{-0.18}$  &
$2.88^{+1.51}_{-1.02}$ & 
$2.18^{+0.31}_{-0.27}$ 
\\

& $kT_{\rm High}$ (keV) & 
$20.5\pm 2.1$ &
$20.9^{+2.7}_{-2.5}$ &
$17.5^{+1.8}_{-1.7}$
\\


& $R_{\rm High}$ (km) &
$0.142^{+0.038}_{-0.028}$ &
$0.0766^{+0.0216}_{-0.0165}$ & 
$0.191^{+0.047}_{-0.0381}$ 
\\

& PIN Flux\footnotemark ($10^{-7}$ erg s$^{-1}$ cm$^{-2}$) & 
$2.54^{+0.12}_{-0.35}$ &
$0.45^{+0.01}_{-0.12}$ &
$1.91^{+0.12}_{-0.29}$
 \\
 
 & Total Flux\footnotemark  ($10^{-7}$ erg s$^{-1}$ cm$^{-2}$) & 
$4.05^{+0.13}_{-0.70}$ &
$0.98\pm 0.02$ &
$3.05^{+0.12}_{-0.45}$
\\

& fit goodness $\chi^2_{\nu}$ ($\nu$) & 
0.62 (27)  &
0.79 (18) &
0.58 (24)
\\

& figure & 
fig.\ref{fig:moderate-intensity_short_burst_spec}(a3,5) & 
fig.\ref{fig:moderate-intensity_short_burst_spec}(b3,5) & 
fig.\ref{fig:moderate-intensity_short_burst_spec}(c3,5)  
\\




 

 
 


\hline 

2BB + PL 
& $kT_{\rm Low}$ (keV) &
$5.85^{+0.67}_{-0.57}$ & 
$2.90^{+0.30}_{-0.28}$ & 
$4.35^{+0.63}_{-0.54}$
\\


& $R_{\rm Low}$ (km) & 
$1.45^{+0.24}_{-0.22}$ & 
$1.99^{+0.37}_{-0.33}$ &
$0.565^{+0.050}_{-0.049}$ 
\\

& $kT_{\rm High}$ (keV) & 
$20.2^{+2.5}_{-2.3}$ & 
$20.9^{+2.9}_{-2.6}$ &
$17.1^{+1.9}_{-1.8}$
\\


& $R_{\rm High}$ (km) &
$0.139^{+0.042}_{-0.033}$ &
$0.0730^{+0.0219}_{-0.0167}$ &
$0.191^{+0.052}_{-0.041}$ 
\\

& $\Gamma_{\rm bst}$ & 
1.54 (fix) & 
1.54 (fix) & 
1.54 (fix)  
\\

& PL Flux\footnotemark ($10^{-7}$ erg s$^{-1}$ cm$^{-2}$) & 
$0.410^{+0.190}_{-0.195}$ & 
$0.140^{+0.092}_{-0.093}$ & 
$0.339^{+0.196}_{-0.200}$
 \\

& fit goodness $\chi^2_{\nu}$ ($\nu$) & 
0.48 (26) & 
0.68 (18) & 
0.49(23)
\\

& figure & 
fig.\ref{fig:2bb_pl.eps}(a) & 
fig.\ref{fig:2bb_pl.eps}(b)  & 
fig.\ref{fig:2bb_pl.eps}(c)  
\\
\hline 
\end{tabular}

\medskip
1:  Fluences are estimated using the CutPL model in the 10--70 keV range with T90. \\
2:  The interstellar absorption (wabs) is multiplied to each model. \\
3, 5, 7:  Absorbed e X-ray fluxes in the 10--70 keV energy range at its average value.  \\
4, 6, 8, 9:  Absorbed  X-ray fluxes in the 1--300 keV energy range at its average value.   \\
\end{flushleft}
\end{table*}

\begin{table*}
\begin{flushleft}
\caption{Spectral parameters of the accumulated weaker short burst.  All the quoted errors are at the 1$\sigma$ level.}
\label{tab: list of spectral fittings of cummulative}
\begin{tabular}{@{}lll}
\hline \hline
& Burst Emission  & \\
\hline
& Summed exposure (sec) & 3.7 \\
& Ave. fluence$^1$ (erg cm$^{-2}$) &$6.61^{+0.04}_{-2.04}\times 10^{-9}$ \\
\hline
Spectral Model & Parameter & Value \\
\hline
PL ($N_{\rm H}$ fixed) & $N_{\rm H}$ ($10^{22}$ cm$^{-2}$) & 3.2 (fix) \\
	& Photon Index $\Gamma_{\rm bst}$ & 1.5 \\
	& fit goodness $\chi^2_{\nu}$ ($\nu$) & 2.05 (34) \\

\hline	
PL & $N_{\rm H}$ ($10^{22}$ cm$^{-2}$) & $5.4^{+0.8}_{-0.5}$ \\ 	
	& Photon Index $\Gamma_{\rm bst}$ & $1.57\pm  0.04$ \\
	& Flux$^{1}$ ($10^{-8}$ erg s$^{-1}$ cm$^{-2}$) & $1.7\pm 0.1$ \\	
	& fit goodness $\chi^2_{\nu}$ ($\nu$) & 1.29 (33) \\	

\hline
CutoffPL & $N_{\rm H}$ ($10^{22}$ cm$^{-2}$) & 3.2 (fix) \\
	& Photon Index $\Gamma_{\rm bst}$ & $1.03\pm0.07$ \\
	& Cutoff Energy $E_{\rm cut}$ (keV) & $62.9^{+14.5}_{-10.8}$ \\
	& Flux$^{1}$ ($10^{-8}$ erg s$^{-1}$ cm$^{-2}$) & $2.3^{+0.1}_{-0.2}$ \\	
	& fit goodness $\chi^2_{\nu}$ ($\nu$) & 0.81 (33) \\	

\hline
2BB & $N_{\rm H}$ ($10^{22}$ cm$^{-2}$) & 3.2 (fix) \\
	& $kT_{\rm Low}$ (keV) & $1.68^{+0.17}_{-0.15}$ \\
	& $R_{\rm Low}$ (km) & $3.33^{+0.52}_{-0.46}$ \\
	& $kT_{\rm High}$ (keV) & $12.3^{+0.9}_{-0.8}$ \\
	& $R_{\rm High}$ (km) & $0.140^{+0.021}_{-0.018}$ \\
	& Flux$^{1}$ ($10^{-8}$ erg s$^{-1}$ cm$^{-2}$) & $2.5\pm 0.2$  \\			
	& fit goodness $\chi^2_{\nu}$ ($\nu$) & 1.63 (32) \\	


\hline
2BB + PL & 	$N_{\rm H}$ ($10^{22}$ cm$^{-2}$) & 3.2 (fix) \\
	& $kT_{\rm Low}$ (keV) & $2.71^{+0.50}_{-0.48}$ \\
	& $R_{\rm Low}$ (km) & $1.16^{+0.41}_{-0.29}$ \\	
	& $kT_{\rm High}$ (keV) & $13.3^{+2.0}_{-1.6}$ \\
	& $R_{\rm High}$ (km) & $0.0905^{+0.0283}_{-0.0255}$ \\	
	& $\Gamma_{\rm h}$ & 1.53 (fix) \\
	& PL Flux$^{1}$ ($10^{-8}$ erg s$^{-1}$ cm$^{-2}$)& $0.78\pm 0.13$ \\
	& fit goodness $\chi^2_{\nu}$ ($\nu$) & 0.82 (31) \\	
	
\hline 	
\hline 
& Persistent Emission$^{2}$  & \\ 
\hline
BB  + PL & $N_{\rm H}$ ($10^{22}$ cm$^{-2}$) & $3.2\pm 0.1$  \\
	& $kT$ (keV) & $0.65^{+0.02}_{-0.01}$ \\
	& $\Gamma_{\rm per}$ & $1.54^{+0.03}_{-0.04}$ \\	
	&  Flux$^{1}$ ($10^{-10}$ erg s$^{-1}$ cm$^{-2}$)  & $1.40^{+0.05}_{-0.07}$  \\
	& fit goodness $\chi^2_{\nu}$ ($\nu$) & 1.08 (278) \\	
\hline 
\end{tabular}

\medskip
1:  Average fluence is evaluated in the 10--70 keV band assuming the 2BB+PL model. Flux are estimated in the 10--70 keV. \\
2:  Values from Paper I. The quoted errors are converted to the 1$\sigma$ level. \\
\end{flushleft}
\end{table*}

\clearpage
\appendix
\section{Correction for the P-sum mode XIS0}
\label{subsection:Timing correction for the P-sum mode}

\begin{figure*}
\begin{center}
\includegraphics[width=160mm]{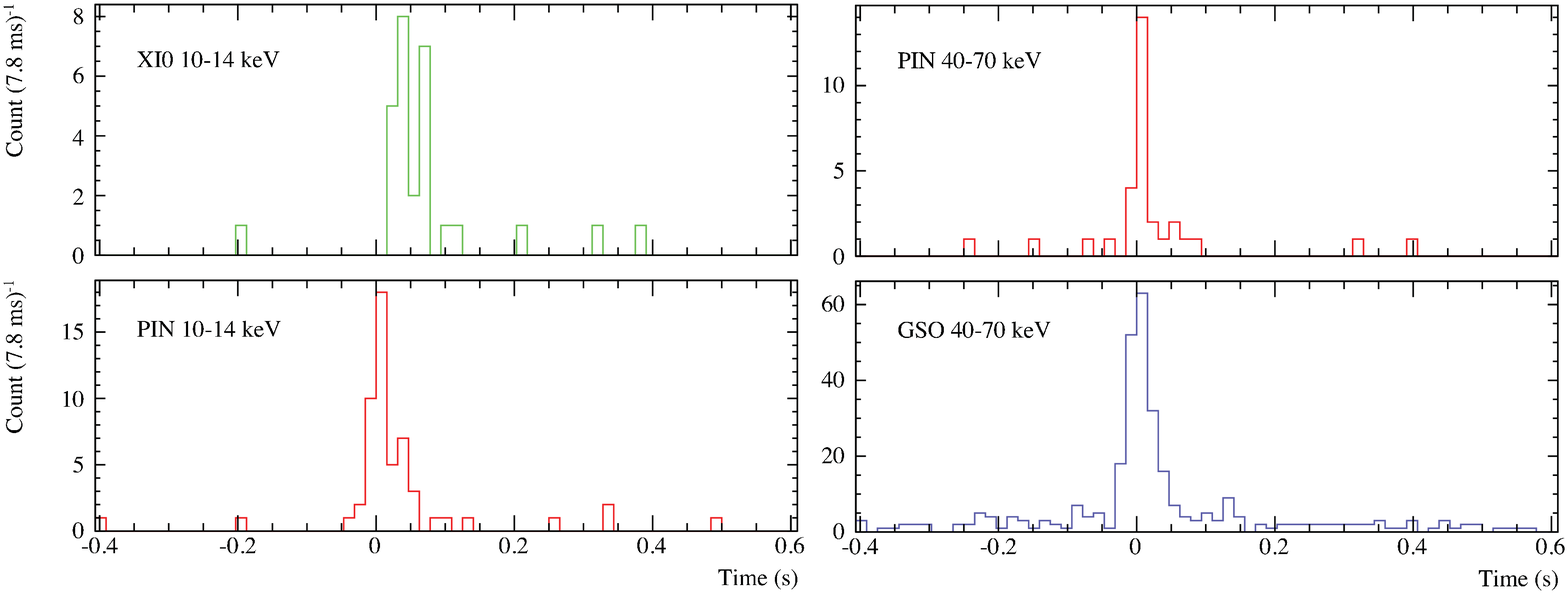}
 \caption{
 (left) 
 Accumulated light curves of 5 short bursts (Burst-4,7, 9, 14, 16) 
	from XIS0 (top) and for HXD-PIN (bottom) both in the 10--14 keV energy range.
(right) 	
Same as the left panels, 
	but from the HXD-PIN (top) and HXD-GSO (bottom)
	both in the 40--70 keV energy range.
Burst-2, 6, 7, 9, 10, 11, 14, 16 and 18 are used for the accumulation.	
}
  \label{fig:accumulated_light_curve}
\end{center}  
\end{figure*}

\begin{figure*}
\begin{center}  
\includegraphics[width=70mm]{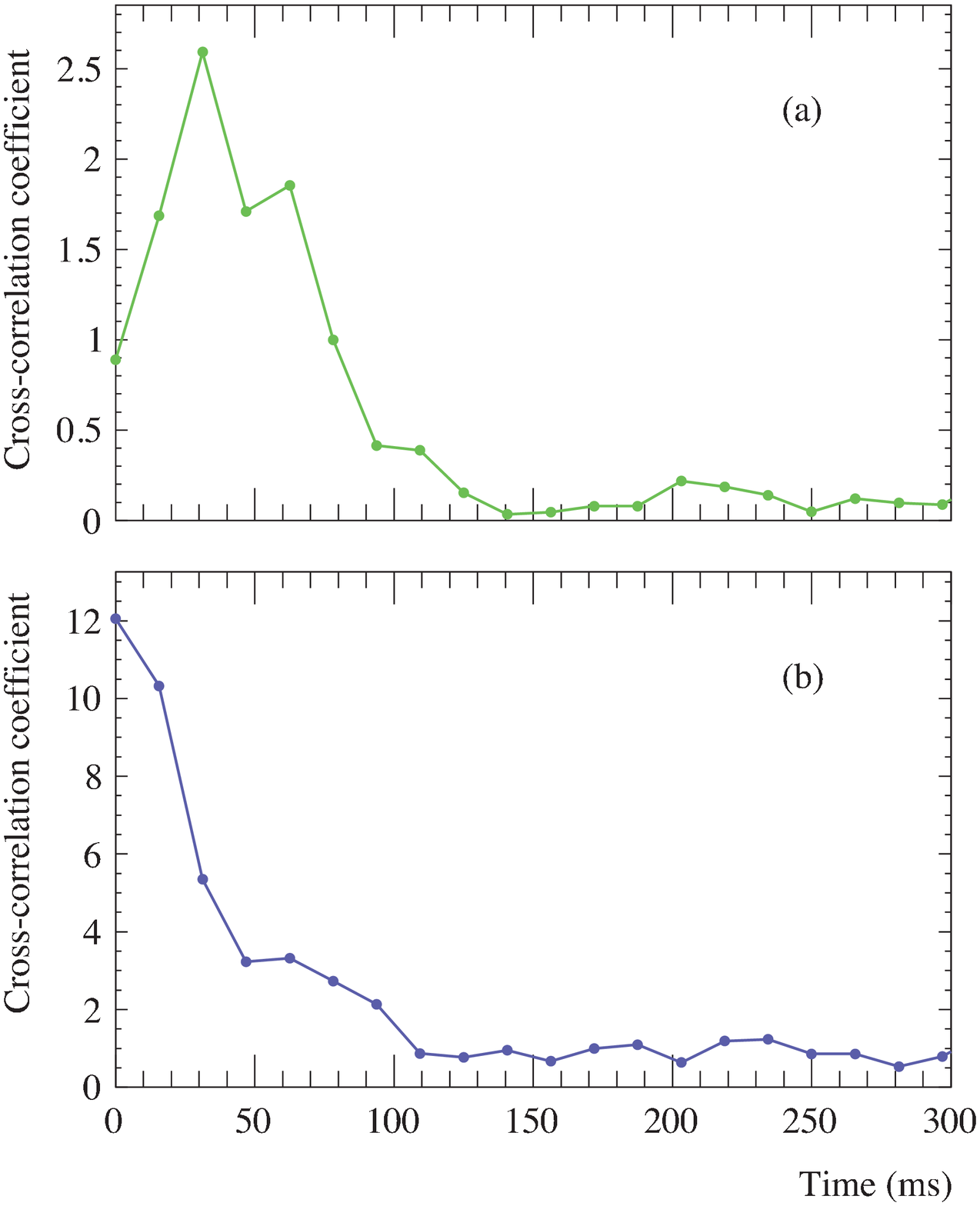}
 \caption{
 Cross-correlation coefficients calculated through equation (\ref{eq:cross-corr}).
Panel (a) is obtained 
	from the P-sum mode XIS0 data to the HXD-PIN data both in the 10--14 keV energy 
  	range, corresponding to the data in figure~\ref{fig:accumulated_light_curve}(left).
Panel (b) is obtained 
	from 	the HXD-GSO data to the HXD-PIN data both in the 40--70 keV energy range,
	corresponding to those in figure~\ref{fig:accumulated_light_curve}(right).	}
  \label{fig:cross_corr}
\end{center}    
\end{figure*}

Here we describe further calibrations of the time assignments and responses of
	the P-sum (timing) mode XIS0, 
	dedicated towards the present analyses.

\subsection{Timing correction}
It is known that
	there is still a slight time lag ($\sim$30 ms) assigned to the P-sum mode of the XIS,
	even after including corrections related with read-out time delays, 
	 depending on the source position on the XIS CCD chip 
	(see \citealt{Matsuta2009SuzakuConf}).
To study the burst light curves in detail,
	we further assigned an additional timing correction to the P-sum mode,
	comparing the HXD-PIN data,
	which was already accurately calibrated using the Crab pulsar \citep{Terada2008PASJ...60S..25T}.
In order to evaluate the residual time lag,
	we compared the burst light curves 
	of the XIS0 to that of HXD-PIN
	both in the 10--14 keV.
We accumulated 5 short bursts
	in which the 10--14 keV photons were clearly detected in the HXD-PIN data.
Panels in Figure~\ref{fig:accumulated_light_curve} (left) 
	are the 10--14 keV cumulative light curves of XIS0 and HXD-PIN, respectively,
	while 
	panels in Figure~\ref{fig:accumulated_light_curve} (right) 
	are 
	the 40--70 keV ones of HXD-PIN and HXD-GSO as a reference.
As compared here,
	the 40--70 keV peaks correspond with each other,
	while
	a peak of the 10--14 keV XIS0 data
	comes slightly later than that of the HXD-PIN.

In order to quantitatively  evaluate this XIS0 time lag,
	we calculated cross-correlations of them.
The cross-correlation coefficience $R_{\rm xy}(i)$ for the $i$-th bin is calculated by
\begin{equation}	
R_{\rm xy}(i) =\frac{1}{N-i}\sum _{j=0}^{N-1-j}x(j)\cdot y(i+j)
\quad
(i=0,1,\cdots,N-1).
\label{eq:cross-corr}
\end{equation}		
Figure~\ref{fig:cross_corr}(a) and (b)
	show  cross-correlation coefficients 
	of the 10--14 keV and 40--70 keV data, respectively.
As clearly shown in these figures,
	no time lag was confirmed between the HXD 40--70 keV data (Figure~\ref{fig:cross_corr}b),
	while 
	the cross-correlation peak of the 10--14 keV range appears at 31.25 ms in Figure~\ref{fig:cross_corr}(a).
Thus, 
	we assigned the time correction of 31.25 ms to the XIS0 data set (\verb+TIME+$-$0.0325 s).
This correction is consistent with the previous analyses 
	of the persistent X-ray emission at \S3.1 in Paper I.
Individual light curves of XIS0 in Figure~\ref{fig:burst_lc_1} 
	were already corrected on this time lag of the 31.25 ms.

\subsection{Response correction}
\label{Response correction}
For the XIS0 spectral analyses,
	we produced 
	rmf and arf files 
	based on the procedure stated in \cite{Matsuta2010Recipi}.
Since the P-sum spectral data 
	is not widely utilized so far,
	it requires a careful calibration,
	we further included correction factors to rmf and arf fiels,
	as follows.
The persistent X-ray emission from 1E~1547.0$-$5408 
	was already measured using the XIS1, XIS3, HXD-PIN, and HXD-GSO data in Paper I.
We also produced the same persistent X-ray spectra from the XIS0,
	after eliminating bright short bursts,
	and 
	compared it with that in Paper I.
We fixed the spectral parameters at the values reported in Paper I,
	and 
	only three parameters of the XIS0 response  
	(a normalization factor, 
	a slope and an offset of the gain)
	were left free to be fitted.
Then,
	we determined 	the normalization factor, the slope and the gain to be 1.079,
	0.90, $-0.05$, respectively.
In this paper, 
	we use these correction factors.

\section{The statistical significances of detected bursts}
\label{subsection:The statistical significances of detected bursts}

We have already evaluated the detection significances in \S2.2 using only the HXD-PIN data.
Here we further evaluated the statistical significance of these detections 
	combing the XIS, HXD-PIN, and HXD-GSO together.
Especially, 
	the XIS instruments are placed at different location in the {\it Suzaku} satellite,
	temporal coincidences of the X-ray events between the HXD and the XIS0
	make the detection much plausible.
A possibility to detect $n$ events,
	from 	the Poisson distribution of its average $\lambda$ count sec$^{-1}$,
	during a time duration of $T$
	can be written as $P(\lambda, n)=(\lambda T)^n \exp (-\lambda T)/n!$.
Thus, multiplied by the independent instruments,
	the total probability of these bursts 
	become $P_{\rm total}=P_{\rm XIS}\cdot P_{\rm PIN} \cdot P_{\rm GSO}$.
Here we employ $T=1$ sec
	and corresponding burst rates listed in Table~\ref{tab: list of burst},
	together with average rate of 
	$\lambda_{\rm XIS0}=3.98$ counts s$^{-1}$,
	$\lambda_{\rm PIN}=0.76$ counts s$^{-1}$,
	$\lambda_{\rm GSO}=13.38$ counts s$^{-1}$ (see \S2.1).
Considering the total bin number of $n_{\rm bin}=4.5\times 10^4$,
	chance occurrence $n_{\rm bin}P_{\rm total}$
	of these bursts are estimated to be below $10^{-5}$.
Therefore,
	we confirmed these events are indeed short bursts.

\section{Check of the data loss}
\label{subsection:Estimation of the data loss}
The burst data (except for Burst-6 and Burst-10),
	were examined for possible losses or pile-ups, 
	through the same procedure as 
	applied to the much brighter burst from SGR~0501+4516 
	recorded with {\it Suzaku} in 2008 \citep{Enoto2009ApJ}.
There are three possible forms of the data loss;
	(1) dead time in the event handling at the analog electronics (HXD-AE),
	(2) the limited event transfer rate from HXD-AE to the digital electronics (HXD-DE),
	and 
	(3) that from HXD-DE to the spacecraft data processor. 
Since the peak count rates of these bursts are less than $\sim$60 counts (15.6 ms)$^{-1}$,
	the effect of (1) is estimated to be at most $\la$8\%. and on average $\sim$1\%,
	since the data acquisition time (20 $\mu$s) is much shorter than
	the burst event interval.
Pileup is much less effective due to a short time window for the pulse height latch.
These event data were not lost in the above (2) and (3) processes as well,
	since the event rates were much smaller than the maximum limits;
	4 kHz for (2) and 400 counts (4 sec)$^{-1}$ for (3).
Thus, 
	our samples are free from
	the data loss due to the dead-time
	or 
	the event pile-ups.
Since the dead-time effect is typically $\sim1$\% for most cases,
	we does not correct the dead-time.

\end{document}